\documentclass[letterpaper,twocolumn,english,aps,prb,floatfix,showpacs]{revtex4}
\usepackage[T1]{fontenc}
\usepackage[latin1]{inputenc}
\usepackage{amsmath}
\usepackage{babel}
\usepackage{graphics}
\usepackage{amssymb}


\begin{document}

\title{Luttinger liquid superlattices: realization of gapless insulating
phases}

\author{J. \surname{Silva-Valencia}}

\email{jeresons@ifi.unicamp.br}

\affiliation{Instituto de Física Gleb Wataghin, Unicamp, C.P. 6165, 13083-970,
Campinas, SP, Brazil}

\author{E. Miranda}

\email{emiranda@ifi.unicamp.br}

\affiliation{Instituto de Física Gleb Wataghin, Unicamp, C.P. 6165, 13083-970,
Campinas, SP, Brazil}

\author{Raimundo R. \surname{dos Santos}}

\email{rrds@if.ufrj.br}

\affiliation{Instituto de Física, Universidade Federal do Rio de Janeiro, C.P.
68528, 21945-970, Rio de Janeiro, RJ, Brazil}

\date{\today}

\begin{abstract}
We investigate LL superlattices, a periodic structure composed of
two kinds of one-dimensional systems of interacting electrons. We
calculate several properties of the low-energy sector: the effective
charge and spin velocities, the compressibility, various correlation
functions, the Landauer conductance and the Drude weight. The low-energy
properties are subsumed into effective parameters, much like homogeneous
one-dimensional systems. A generic result is the \emph{weighted average}
\emph{nature} \emph{of these parameters,} \emph{in proportion to the
spatial extent} \emph{of the underlying subunits}, pointing to the
possibility of {}``engineered{}'' structures. As a specific realization,
we consider a one-dimensional Hubbard superlattice, which consists
of a periodic arrangement of two long Hubbard chains with different
coupling constants and different hopping amplitudes. This system exhibits a
rich phase diagram with several phases, both metallic and insulating.
We have found that gapless insulating phases are present over a wide range 
of parameters.
\end{abstract}

\maketitle

\section{INTRODUCTION}

The physics of one-dimensional electronic systems has been the subject
of a vigorous onslaught recently, both theoretical and experimental.
Experimentally, the ability to grow nanostructures such as quantum
wires\cite{tarucha, yacobyetal, fazioetal} and carbon 
nanotubes\cite{iijima, endoetal, ebbesen, tansetal, bockrathetal, hamadaetal}
has enabled, for the first time, the investigation of systems of a
truly one-dimensional nature. On the theoretical side, the peculiarities
of the behavior of interacting electrons in one dimension have culminated
in the proposal of a unique universality class dubbed the Luttinger
liquid\cite{luttinger1, luttinger2, lutherpeschel1, solyom, Emery, haldaneLL, voit}
(LL), which stands in sharp contrast with the higher dimensional Fermi
liquids established by Landau. The LL is characterized by the absence
of stable quasi-particles, its low-energy sector being exhausted by
collective charge and spin density excitations. Since the latter travel
at different velocities, an added electron splits up into well separated
charge and spin degrees of freedom. Furthermore, correlation functions
decay in a power law fashion, with exponents set by only a few parameters.
This generic behavior has been tested and confirmed in the case of
edge transport in systems which exhibit the fractional quantum Hall
effect.\cite{kanefisher2, moon, milliken, changetal, graysonetal}
LL theory has also been successfully used to describe some low-energy
properties of carbon nanotubes,\cite{bockrathetal2, Yaoetal, eggeretal}
though the situation in quantum wires remains
controversial.\cite{auslaenderetal, wangetal}

The effect of boundary conditions on the low-energy properties
of LL's was first considered several years ago.\cite{fabriziogogolin}
Moreover, the interplay between boundary, finite-size, and thermal
effects has been shown to alter considerably the properties of the
system.\cite{eggertetal, mattson} In particular, the zero-temperature
critical behavior of the bulk always crosses over to a boundary dominated
regime. These studies are important to explain the experimental results
of tunneling spectroscopy into one-dimensional systems. More recently,
it has been proposed that one-dimensional systems with gapless degrees
of freedom and open boundary conditions form a new universality class
of quantum critical behavior called `bounded Luttinger
liquids'.\cite{voitwanggrioni} 

A particular kind of boundary effect emerges in the case of
inhomogeneities. In general, an inhomogeneous LL is modeled by
allowing the velocities of collective excitations \( u_{\rho } \) and
\( u_{\sigma } \) and the correlation exponents \( K_{\rho } \) and \(
K_{\sigma } \) to vary in space. The absence of conductance
renormalization in long high-mobility GaAs wires, for instance, has
been analyzed and explained in terms of an inhomogeneous LL model,
where the Fermi liquid leads are replaced by a non-interacting
one-dimensional electron gas.\cite{maslovstone, safischulz1,
  ponomarenko, safischulz2, ponomarenkonagaosa} Furthermore, LL's with
different inhomogeneity profiles have also been used in the context of
the fractional quantum Hall effect, to describe transitions between
edge states at different fillings,\cite{oregfinkelstein,
  chklovskiihalperin} or between an edge state and a Fermi
liquid.\cite{chamonfradkin}

With an eye to practical applications as diodes or transistors,
researchers have recently begun to fabricate heterojunctions of carbon
nanotubes\cite{chico, collinsetal, marteletal, tansetal2, Yaoetal,
  kilicetal, andriotisetal} which look especially promising. They
happen to be another realization of an inhomogeneous one-dimensional
system. Taking this idea one step further, we have been led to
consider another kind of heterostructure: \emph{a superlattice}. The
effect of electronic correlations in superlattices was initiated
through a one-dimensional Hubbard-like model called a Hubbard
superlattice (HSL), \cite{paivasantos1, paivasantos2, paivasantos4}
consisting of a periodic arrangement where the Hubbard on-site
repulsion \( U \) is turned on and off in a repeated fashion.  Despite
its simplicity, a number of remarkable features were found, in marked
contrast with the otherwise homogeneous system: local moment weight
can be transferred from repulsive to free sites, spin density wave
(SDW) quasi-order is wiped out as a result of frustration, and strong
SDW correlations (in a subset of sites) could set in above
half-filling.  Furthermore, the evolution of the local moment and of
the charge gap, together with a strong-coupling analysis, showed that
the electron density at which the system becomes a Mott insulator
increases with the size of the free layer relative to the repulsive
one. More recently, the possibility of a periodically modulated
hopping at arbitrary filling and magnetization has been
considered.\cite{cabraetal2}

In order to generalize the effects of a superlattice structure in an
interacting one-dimensional system, we consider here a general
Luttinger liquid superlattice (LLSL), making at first no reference to
the underlying microscopic details. We show how its low-energy
properties bear strong resemblance to a conventional Luttinger liquid.
However, as in the case of bounded Luttinger
liquids,\cite{voitwanggrioni} new effective parameters have to be
introduced, which are the superlattice analogues of the spin and
charge velocities and stiffnesses. These encode all the information
necessary for a description of the low-energy sector. Moreover, these
effective parameters turn out to mix the properties of the underlying
sub-units in proportion to their spatial extent.  This spatial
averaging characteristic suggests the possibility of fine-tuning the
physical properties by a careful selection of the superlattice
modulation, a feature which may prove useful in nano-device
applications. We then consider specific realizations of the LLSL by
analyzing in full detail a general HSL. We find a proliferation of
phases, both metallic and insulating.  Surprisingly, the insulating
phases often have no charge gap, because additional charge can be
accommodated in the compressible sub-units.  A partial account of
these results has appeared in Ref.\ \onlinecite{valencia1}.

The paper is organized as follows: In Sec. \ref{model}, we introduce
the bosonic formulation of the Tomonaga-Luttinger model and our model.
We obtain the effective charge and spin velocities, the correlation
functions with the effective exponents and the Drude weight for LL
superlattices. The application of these results to various cases where
the LL describes the low-energy sector of a Hubbard model is analyzed
in Sec. \ref{hubsl}. We close with the conclusions in
Sec. \ref{conclusions}.

\section{THE MODEL}

\label{model}We briefly review the general aspects of a homogeneous
LL in order to set up the notation. The low-energy, large-distance
behavior of a one-dimensional fermionic system with spin-independent
interactions is described by the Hamiltonian\cite{luttinger1, luttinger2, 
lutherpeschel1, solyom, Emery, haldaneLL, voit}
\begin{equation}
\label{HLL}
H=H_{\rho }+H_{\sigma }+\frac{2g_{1}}{(2\pi \alpha )^{2}}\int dx\cos
(\sqrt{8}\Phi _{\sigma }),
\end{equation}
where $\alpha$ is a short-distance cutoff, \(g_{1}\) is the spin
backward-scattering amplitude, and
\begin{equation}
\label{Hnu}
H_{\nu }=\int dx\ \left(\frac{\pi u_{\nu }K_{\nu }}{2}\Pi ^{2}_{\nu
}+\frac{u_{\nu }}{2\pi K_{\nu }}(\partial _{x}\Phi _{\nu })^{2}\right),
\end{equation}
with $\nu=\rho$ and $\sigma$ for the charge and spin degrees of freedom,
respectively.

The phase fields are
\begin{equation}
\label{phil}
\Phi _{\nu }(x)=
-\frac{i\pi }{L}\sum _{p\neq 0}\frac{1}{p}e^{-\alpha \left| p\right| x/2-ipx}
[\nu _{+}(p)+\nu _{-}(p)]-N_{\nu }\frac{\pi x}{L},
\end{equation}
and
\begin{equation}
\label{pil}
\Pi _{\nu }(x)=\frac{1}{L}\sum _{p\neq 0}e^{-\alpha \left| p\right| x/2-ipx}[\nu _{+}(p)-\nu _{-}(p)]+\frac{J_{\nu }}{L}.
\end{equation}
Here \( \rho _{r}(p)\,[\sigma _{r}(p)] \) are the Fourier components
of the charge- (spin-) density operator for the right- ($r=+$) and 
left- ($r=-$) branches of moving fermions. Introducing the total number
operators (measured with respect to the ground state) \( N_{rs} \)
for branch \( r \) and spin \( s \), the total (charge and spin)
number and current operators \( N_{\nu },J_{\nu } \) are 
\begin{equation}
\label{Nu}
N_{\nu }=\frac{1}{\sqrt{2}}[(N_{+,\uparrow }+N_{-,\uparrow })\pm (N_{+,\downarrow }+N_{-,\downarrow })],
\end{equation}
and
\begin{equation}
\label{Ju}
J_{\nu }=\frac{1}{\sqrt{2}}[(N_{+,\uparrow }-N_{-,\uparrow })\pm (N_{+,\downarrow }-N_{-,\downarrow })],
\end{equation}
where the upper and lower signs correspond to \( \nu =\rho  \) and
\( \sigma  \), respectively.

The operators \( \Phi _{\nu } \) and \( \Pi _{\nu } \) in Eqs.\ 
(\ref{HLL}) and (\ref{Hnu}) obey Bose-like commutation relations: 
\( [\Phi_{\nu }(x),\Pi _{\mu }(y)]=i\delta _{\nu \mu }\delta (x-y) \).
Consequently, at least for \( g_{1}=0 \), Eq.\ (\ref{HLL}) describes
independent long-wavelength oscillations of the charge and spin density,
with linear dispersion relations \( \omega _{\nu }(k)=u_{\nu }\left| k\right|  \),
(\( u_{\nu } \) is the velocity of elementary excitations) and the
system is conducting. The only nontrivial interaction effects in (\ref{HLL})
come from the cosine term. However, for repulsive SU(2) invariant
interactions (\( g_{1}>0 \)), this term is renormalized to zero in
the long-wavelength limit, and at the fixed point one has \( K^{*}_{\sigma }=1 \).
The three remaining parameters in (\ref{HLL}) then completely determine
the long-distance properties of the system; in particular, \( K_{\rho
} \)
determines the long-distance decay of all the correlation functions
of the system.
\begin{figure}
{\centering \resizebox*{3.4in}{!}{\includegraphics*{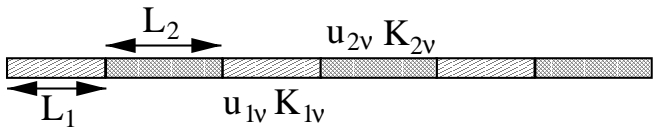}} \par}
\caption{\label{flls}Schematic representation of a Luttinger liquid
superlattice. Here, \protect\( u_{\lambda ,\nu }\protect \), \protect\(
K_{\lambda ,\nu }\protect \) and \protect\( L_{\lambda }\protect \) are
the velocities, interaction parameters and sizes of two Luttinger liquids
(\protect\( \lambda \protect \)=1,2).} 
\end{figure} 

We now consider a LLSL, consisting of a repeated
pattern of two different LL's with parameters \( u_{\lambda ,\nu } \),
\( K_{\lambda ,\nu } \) and sizes \( L_{\lambda } \) (\( \lambda =1,2 \))
perfectly connected (Fig.\ \ref{flls}). We use the adiabatic 
approximation,
in which the scale of the inhomogeneity between the two liquids is
much larger than the Fermi wavelength \( 2\pi /k_{F} \). Thus, the
single-particle backscattering from the inhomogeneities can be neglected.
Accordingly, the low-energy properties of this LLSL
are described by generalizing the usual bosonized Hamiltonian of
Eq.~(\ref{HLL}) as follows: 

\begin{eqnarray}
H & = & \frac{1}{2\pi }\sum _{\nu =\rho ,\sigma }\int dx
\ \biggl\{ u_{\nu }(x)K_{\nu }(x)\left( \partial _{x}\Theta _{\nu
}\right)^{2} 
\nonumber \\
 & + &  \frac{u_{\nu }(x)}{K_{\nu }(x)}\left( \partial _{x}\Phi
_{\nu }\right)^2
\biggr\},
\label{llsl} 
\end{eqnarray}
where the sum extends over separated charge- (\( \nu =\rho  \))
and spin- (\( \nu =\sigma  \)) degrees of freedom, each of which
with interaction- and layer-dependent parameters \( u_{\nu }(x) \)
and \( K_{\nu }(x) \). For \( x \) on the first (second) `layer'
one has \( K_{\nu }(x)=K_{1,\nu } \)(\( K_{2,\nu } \)) and \( u_{\nu }(x)=u_{1,\nu } \)
(\( u_{2,\nu } \)).

The boson phase fields \( \Phi _{\nu } \) are related to the charge
and spin densities, \( \rho  \) and \( \sigma  \), through \( \sqrt{2}\partial _{x}\Phi _{\nu }(x)/\pi =\nu  \),
while \( \Theta _{\nu } \) is such that \( \partial _{x}\Theta _{\nu } \)
is the momentum field conjugate to \( \Phi _{\nu } \): \( [\Phi _{\nu }(x),\partial _{y}\Theta _{\mu }(y)]=i\delta _{\nu ,\mu }\delta (x-y) \).
Note that $\Pi_{\nu}(x)=\partial_x\Theta_{\nu}(x)$ in Eqs.  
(\ref{HLL}) and (\ref{pil}).

The equations of motion for the fields \( \Phi _{\nu } \) and \( \Theta _{\nu } \)
are
\begin{eqnarray}
\partial _{t}\Phi _{\nu } & = & u_{\nu }(x)K_{\nu }(x)\partial _{x}\Theta _{\nu },\label{em1} \\
\partial _{t}\Theta _{\nu } & = & \frac{u_{\nu }(x)}{K_{\nu }(x)}\partial
_{x}\Phi_{\nu },\label{em2} 
\end{eqnarray}
which illustrate their duality under the replacement \( K_{\nu }\left( x\right) \rightarrow 1/K_{\nu }\left( x\right)  \).
Substituting (\ref{em2}) into (\ref{em1}) yields \begin{equation}
\label{emphi}
\partial _{tt}\Phi _{\nu }-u_{\nu }K_{\nu }\partial _{x}\left( \frac{u_{\nu }}{K_{\nu }}\partial _{x}\Phi _{\nu }\right) =0,
\end{equation}
 and a similar equation for \( \Theta _{\nu } \).

We now have to set up the matching equations at the interfaces between
layers. The equations of motion lead to the continuity of \( \Phi _{\nu } \)
and \( \Theta _{\nu } \) and their time derivatives. The right hand
sides of Eqs.\ (\ref{em1}-\ref{em2}) yield, as additional conditions,
the continuity of both \( \left( u_{\nu }/K_{\nu }\right) \partial
_{x}\Phi _{\nu } \)
and \( u_{\nu }K_{\nu }\partial _{x}\Theta _{\nu } \) at the contacts.
Note that the continuity of \( \Phi _{\nu } \) and \( \Theta _{\nu } \)
guarantees that of the fermionic field.\cite{safischulz1, maslovstone,
ponomarenko}
Physically, these boundary conditions simply encode \emph{the conservation
of both charge and spin currents} \( j_{\nu }=\sqrt{2}\partial _{t}\Phi _{\nu }/\pi  \)
(since we are neglecting Umklapp processes and backscattering of electrons
with opposite spin). We stress that, under these conditions, these
are the only universal requirements on the fields, irrespective of
the actual interface potentials.

The superlattice structure is incorporated into the solution of the
equations of motion in a way completely analogous to the discussion
of reflection and transmission in the Kronig-Penney model. That is,
we diagonalize the Hamiltonian (\ref{llsl}) by expanding the phase
fields in normal modes
\begin{widetext}
\begin{eqnarray}
\Phi _{\nu }(x,t) & = & -i\sum _{p\neq 0}{\textrm{sign}}(p)\frac{\phi _{p,\nu }(x)}{2\sqrt{\omega _{p,\nu }}}\left[ b_{-p,\nu }e^{i\omega _{p,\nu }t}+b_{p,\nu }^{\dagger }e^{-i\omega _{p,\nu }t}\right]
 - \phi _{0,\nu }(x)+\gamma _{\lambda \nu }t,\label{phis} \\
\Theta _{\nu }(x,t) & = & i\sum _{p\neq 0}\frac{\theta _{p,\nu }(x)}{2\sqrt{\omega _{p,\nu }}}\left[ b_{-p,\nu }e^{i\omega _{p,\nu }t}-b_{p,\nu }^{\dagger }e^{-i\omega _{p,\nu }t}\right] 
  +  \theta _{0,\nu }(x)-\tau _{\lambda \nu }t,
\end{eqnarray}
\end{widetext}
where \( b_{p,\nu }^{\dagger } \), are boson creation operators (\( p>0 \)).
The normal mode eigenfunctions \( \phi _{p,\nu }(x) \) and eigenvalues
\( \omega _{p,\nu } \) satisfy \begin{equation}
\label{eigen}
\omega _{p,\nu }^{2}\phi _{p,\nu }(x)+u_{\nu }K_{\nu }\partial _{x}\left( \frac{u_{\nu }}{K_{\nu }}\partial _{x}\phi _{p,\nu }\right) =0,
\end{equation}
 {[}obtained by taking (\ref{phis}) into (\ref{emphi}){]}, subject
to the same boundary conditions at the contacts as before, with \( \phi _{p,\nu }(x) \)
replacing \( \Phi _{\nu }(x) \). The eigenvalues are given by

\begin{eqnarray}
\cos p(L_{1}+L_{2}) & = & \cos (\frac{\omega _{p,\nu }L_{2}}{u_{2,\nu }})\cos (\frac{\omega _{p,\nu }L_{1}}{u_{1,\nu }})\nonumber \\
 & - & \frac{\Delta _{\nu }}{2}\sin (\frac{\omega _{p,\nu }L_{2}}{u_{2,\nu }})\sin (\frac{\omega _{p,\nu }L_{1}}{u_{1,\nu }}),
\label{espectro}
\end{eqnarray}
where \( \Delta _{\nu }=\eta _{\nu }+\eta ^{-1}_{\nu } \) and \( \eta
_{\nu }=K_{1,\nu }/K_{2,\nu } \).
For \( p\ll \pi /(L_{1}+L_{2}) \), the dispersion relation of the
LLSL is linear, i.e., \( \omega _{\nu }(p)=c_{\nu }\left| p\right|  \),
with an effective velocity 
\begin{equation}
\label{velo}
c_{\nu }=\frac{u_{1,\nu }(1+\ell )}{\sqrt{1+\Delta _{\nu }\ell u_{1,\nu }/u_{2,\nu }+\left( \ell u_{1,\nu }/u_{2,\nu }\right) ^{2}}},
\end{equation}
where \( \ell \equiv L_{2}/L_{1} \); clearly, \( c_{\nu }\rightarrow
u_{2,\nu } \) as \( \ell \rightarrow \infty \), and \( c_{\nu
  }\rightarrow u_{1,\nu } \) as \( \ell \rightarrow 0 \).  Also, from
Eq. (\ref{espectro}) it follows that the spectrum of elementary
excitations of a LLSL has bands and gaps, reflecting the superlattice
structure. In this regard, it should be mentioned that, for a Luttinger
liquid with a periodically modulated particle density, the presence of
a plasmon gap was reported.\cite{gramada} Here, we will focus only on
the low energy properties of the LLSL.

On the other hand, the zero mode functions \( \phi _{0,\nu }(x) \)
and \( \theta _{0,\nu }(x) \), satisfy \begin{eqnarray}
\tau _{\lambda \nu } & = & \frac{u_{\nu }(x)}{K_{\nu }(x)}\partial _{x}\phi _{0,\nu }(x),\\
\gamma _{\lambda \nu } & = & u_{\nu }(x)K_{\nu }(x)\partial _{x}\theta _{0,\nu }(x),\label{equzer} 
\end{eqnarray}
 which follow from Eqs. (\ref{em1}) and (\ref{em2}). 
While for the homogeneous system one has
\begin{equation}
\phi _{0,\nu }\left( x\right)  = \pi \frac{N_{\nu}}{L}x,
\label{homphi}
\end{equation}
and
\begin{equation}
\theta _{0,\nu }\left( x\right) = \pi \frac{J_{\nu }}{L}x,
\label{homtheta} 
\end{equation}
for the LLSL there will be, in general, an inhomogeneous periodic
density profile. As we will see, there is a tendency for the charge
to accumulate more in the \emph{less interactive layer.} Thus, the
zero mode functions will reflect this inhomogeneity.\cite{castronetoetal}
Now, since each layer is a LL, \( \phi _{0,\nu } \) and \( \theta _{0,\nu}
\)
will vary in such a way that \( \Delta \phi _{0,\nu }=\pi N_{\lambda \nu } \)
and \( \Delta \theta _{0,\nu }=\pi J_{\lambda \nu } \) across each
layer \( \lambda  \), with layer-specific number and current operators.
We then obtain 
\begin{eqnarray}
\phi _{0,\nu }(x) & = & A_{m,\lambda \nu }+\frac{\pi N_{\lambda \nu }x}{L_{\lambda }},\label{phi0} \\
\theta _{0,\nu }(x) & = & B_{m,\lambda \nu }+\frac{\pi J_{\lambda \nu }x}{L_{\lambda }},\label{theta0} 
\end{eqnarray}
 where
\begin{equation}
\label{amln}
A_{m,\lambda \nu }=\left\{ \begin{array}{c}
\left( m-1\right) \pi L_{2}\left( \frac{N_{2\nu }}{L_{2}}-\frac{N_{1\nu }}{L_{1}}\right) \textrm{ if }\lambda =1,\\
m\pi L_{1}\left( \frac{N_{1\nu }}{L_{1}}-\frac{N_{2\nu }}{L_{2}}\right) \textrm{ if }\lambda =2,
\end{array}\right. 
\end{equation}
 with an analogous expression for \( B_{m,\lambda ,\nu } \) obtained
with the replacement of \( N_{\lambda ,\nu } \) by \( J_{\lambda ,\nu } \).
Here \( m=1,2,3,.... \) labels the unit cell. Analogously, from
Eqs.~(\ref{em1}) and (\ref{em2}) we have
\begin{eqnarray}
\gamma _{\lambda \nu } & = & \pi u_{\nu }(x)K_{\nu }(x)\frac{J_{\lambda \nu }}{L_{\lambda }},\label{gamma} \\
\tau _{\lambda \nu } & = & \pi \frac{u_{\nu }(x)}{K_{\nu }(x)}\frac{N_{\lambda \nu }}{L_{\lambda }}.\label{tau} 
\end{eqnarray}
In a LL, the ground state value of \( \tau _{\rho } \) measures the
charge compressibility, whereas \( \tau _{\sigma } \) is related to
the spin susceptibility. Considering the LLSL zero modes
[Eqs.\ (\ref{phi0}) and (\ref{theta0})] and the Hamiltonian
(\ref{llsl}) we find that
the superlattice compressibility is given by 
\begin{equation}
\label{comsu}
\frac{1}{\kappa _{s}}=\frac{1+\ell }{\kappa _{1}+\ell \kappa _{2}},
\end{equation}
 where \( \kappa _{\lambda }=2K_{\lambda ,\rho }/\pi u_{\lambda ,\rho } \)
is the compressibility of each layer. Clearly \( \kappa _{s} \) is
nothing but an average of the individual compressibilities weighted
by the layer lengths.

Interactions in a one-dimensional system can enhance charge density or
superconducting fluctuations depending on whether they are repulsive
or attractive. Let us then consider the correlation functions for the
LLSL at $T=0$. The asymptotic (i.e., for well separated \( x \) and \(
y \)) behavior of the density-density correlation function is
\begin{eqnarray}
\left\langle n(x)n(y)\right\rangle & \sim & \frac{\alpha _{\rho }}{\pi
^{2}\left| x-y\right| ^{2}}+A_{1}\frac{e^{2i\left( \overline{\phi
}(x)-\overline{\phi }(y)\right) }}{\left| x-y\right| ^{K_{\rho }^{\ast
}+K_{\sigma }^{\ast }}}\nonumber \label{fcdd}\\
& + & A_{2}\frac{e^{4i\left( \overline{\phi }(x)-\overline{\phi
}(y)\right) }}{\left| x-y\right| ^{4K_{\rho }^{\ast
}}},
\label{dens-dens-corr}
\end{eqnarray}
where 
\begin{eqnarray}
\label{knustar}
K_{\nu }^{\ast }&=&\frac{\sqrt{1+\Delta _{\nu }\ell u_{1,\nu }/u_{2,\nu
}+\left( \ell u_{1,\nu }/u_{2,\nu }\right) ^{2}}}{\frac{1}{K_{1,\nu
}}+\ell \frac{1}{K_{2,\nu }}\frac{u_{1,\nu }}{u_{2,\nu }}}\nonumber \\
&\equiv& f(K_{1,\nu },K_{2,\nu }),
\end{eqnarray} 
\begin{equation}
\label{alphanu}
\alpha _{\nu }=K_{\nu }^{\ast }\left( \frac{1+\ell }{\frac{K_{1,\nu
}}{K_{2,\nu }}+\ell \frac{u_{1,\nu }}{u_{2,\nu
}}}\right) ^{2}\times\left\{ \begin{array}{cc}
\left( \frac{K_{1,\nu }}{K_{2,\nu }}\right) ^{2} & \textrm{if }x\textrm{ and }y\in 1,\textrm{ }\\
\frac{K_{1,\nu }u_{1\nu }}{K_{2,\nu }u_{2\nu }} & \textrm{if }(x,y)\in (1,2),\\
\left( \frac{u_{1\nu }}{u_{2\nu }}\right) ^{2} & \textrm{if }x\textrm{ and }y\in 2,
\end{array}\right. 
\end{equation}
and \( \overline{\phi }(x)=k_{F}x-\phi _{0,\rho }(x) \).
The second and third terms on the right-hand side of
Eq.~(\ref{dens-dens-corr}) respectively correspond to the \(2k_{F} \) and
\( 4k_{F} \) correlations in the homogeneous case. And, similarly to the
homogeneous system, the former dominate over the latter for \(K_{\rho
}^{\ast }\geq \frac{1}{3} \) (see, however, Ref.\
\onlinecite{paivasantos3}).

The correlation functions for spin-spin, singlet (SS) and triplet
(TS) superconducting pairing are given by 
\begin{eqnarray}
\left\langle {\textbf {S}}(x).{\textbf {S}}(y)\right\rangle  & \sim  & \frac{\alpha _{\sigma }}{\pi ^{2}\left| x-y\right| ^{2}}+B_{1}\frac{e^{2i\left( \overline{\phi }(x)-\overline{\phi }(y)\right) }}{\left| x-y\right| ^{K_{\rho }^{\ast }+\overline{K}_{\sigma }}}\nonumber \label{fcss} \\
 & + & B_{2}\frac{e^{2i\left( \overline{\phi }(x)-\overline{\phi }(y)\right) }}{\left| x-y\right| ^{K_{\rho }^{\ast }+K_{\sigma }^{\ast }}},\label{spin-spin-corr} \\
\left\langle O_{SS}^{\dagger }(x)O_{SS}(y)\right\rangle  & = & \left\langle O_{TS_{0}}^{\dagger }(x)O_{TS_{0}}(y)\right\rangle \\
 & \sim  & \frac{C_{1}}{\left| x-y\right| ^{\overline{K}_{\rho }+K_{\sigma }^{\ast }}},\\
\left\langle O_{TS_{\pm 1}}^{\dagger }(x)O_{TS_{\pm 1}}(y)\right\rangle  & \sim  & \frac{C_{2}}{\left| x-y\right| ^{\overline{K}_{\rho }+\overline{K}_{\sigma }}},
\end{eqnarray}
where \( \overline{K}_{\nu }=f(1/K_{1,\nu },1/K_{2,\nu }) \) [Eq.\ 
(\ref{knustar})], reflecting the duality properties (in the
homogeneous limit we have \( \overline{K}_{\nu }\rightarrow 1/K_{\nu }
\)). One should note that the correlation functions depend not only on
the difference \( x-y \), but also on the actual positions \( x \) and
\( y \), through the zero mode functions. It is interesting to note
that, even though we now have new effective coupling constants (\(
K_{\nu }^{\ast } \),\( \overline{K}_{\nu } \)), the scaling laws
between the exponents of the correlation functions are not broken by
the superlattice structure. In other words, the replacement
$K_\nu \rightarrow K^\ast_\nu$ and $K_\nu^{-1}\rightarrow
\overline{K}_\nu$ in the exponents of the correlation functions of the
homogeneous system yields the exponents given above for the
superlattice. 

Finally, we discuss the conducting properties. Let us first
consider a LLSL in the presence of a weak external space- and time-dependent
electrostatic potential \( V(x,t) \), such that the electric field
\( E(x,t)=-\partial _{x}V(x,t) \). The interaction of the fermions
with \( V(x,t) \) is described by a source term \begin{equation}
\label{Hext}
H_{ext}=-e\int dx\rho (x)V(x,t).
\end{equation}
Now the equation of motion for \( \Phi _{\rho } \) is\cite{safischulz1, safischulz2, maslovstone, ponomarenko}\begin{equation}
\label{emelec}
\left[ -\frac{\partial _{tt}}{u_{\rho }\left( x\right) K_{\rho }\left( x\right) }+\partial _{x}\left( \frac{u_{\rho }\left( x\right) }{K_{\rho }\left( x\right) }\partial _{x}\right) \right] \Phi _{\rho }(x,t)=-eE(x,t).
\end{equation}
Defining the bosonic Green's function 
\begin{equation}
G(x,y,t)=-i\theta (t)\left\langle \left[ \Phi _{\rho }(x,t),\Phi _{\rho }(y,0)\right] \right\rangle ,
\end{equation}
 the nonlocal conductivity is given by 
\begin{equation}
\label{sigma}
\sigma (x,y,t)=-\frac{2g_{0}}{\pi }\partial _{t}G(x,y,t),
\end{equation}
where \( g_{0}=e^{2}/h \) is the conductance quantum. First, we consider
the usual order of limits, taking \( q\rightarrow 0 \) before \( \omega \rightarrow 0, \)
which yields the Drude weight, appropriate for a situation of a uniform
static electric field.\cite{fenton} 
In this case
\begin{equation}
\label{druw}
\sigma (q=0,\omega \rightarrow 0)=2g_{0}c_{\rho }K_{\rho }^{\ast }\delta
(\omega ),
\end{equation}
which has the same form as for the homogeneous case,\cite{schulz1}
but with the effective velocity and effective exponent replacing the
corresponding uniform quantities \( u_{\rho } \) and \( K_{\rho } \).
Taking the limits in the reverse order yields the Landauer conductance,
which corresponds to a situation where an electric field is applied
to a finite region of the sample.\cite{fenton} In the LLSL we have
\begin{equation}
\label{landauer}
\sigma (q\rightarrow 0,\omega=0)=2g_{0}K^{\ast }_{\rho }\delta \left(
q\right),
\end{equation}
 which is similar to the homogeneous case,\cite{apelrice} except
that the effective exponent appears. Naturally, the conductance renormalization
of Eq.\ (\ref{landauer}) is usually hidden in the presence of Fermi
liquid leads.\cite{maslovstone, ponomarenko, safischulz1} However,
it should be accessible in AC measurements, if \( \omega >c_{\rho }/L \),
the inverse traversal time of the sample.\cite{matveevglazman}

\begin{figure}
{\centering \resizebox*{3.4in}{!}{\includegraphics*{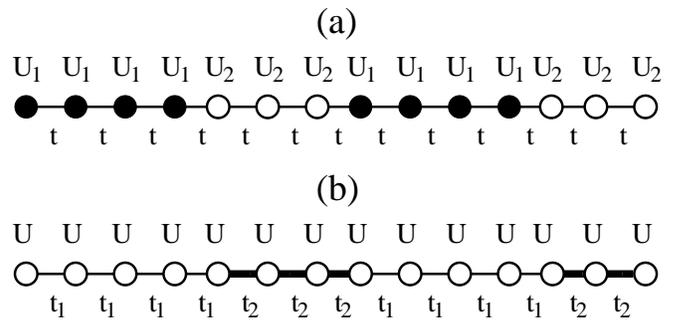}} \par}
\caption{Schematic representation of Hubbard superlattices. In (a)
the hopping is uniform but the interaction is \protect\( U_{\lambda}\protect\)
in the \protect\(\lambda\protect\) sub-chain. In (b) the
interaction is uniform, whereas the hopping can assume two values.
}
\label{slscheme}
\end{figure} 

\section{Hubbard Superlattices}
\label{hubsl}

For the sake of illustrating the LLSL with a specific
realization, we now discuss a one-dimensional Hubbard superlattice
(HSL).\cite{paivasantos1, paivasantos2, paivasantos4, cabraetal2} 
We first consider a periodic arrangement of \( L_{1} \) sites in which
the on-site coupling is \( U_{1}\geq 0 \), followed by \( L_{2} \) others
with on-site coupling \( U_{2}>0 \); the hopping parameter, \( t \), is
uniform, as shown in Fig.\ \ref{slscheme}(a).  We
subsequently consider the on-site interaction as being uniform but the
hopping integrals as periodic: $t_1$ between $L_1$ sites, followed by
\( t_{2}\) between $L_2$ sites; see Fig.\ \ref{slscheme}(b).

Both cases above are contemplated if one writes the Hamiltonian as 
\begin{equation}
\label{hus}
H=-\sum _{i,\sigma }t_{i,i+1}(c^{\dagger }_{i\sigma }c_{i+1\sigma
}+H.c.)+\sum _{i}U_{i}\, n_{i\uparrow }n_{i\downarrow },
\end{equation}
 where, in standard notation, \( i \) runs over the sites of a one-dimensional
lattice, \( c^{\dagger }_{i\sigma }(c_{i\sigma }) \) creates (annihilates)
a fermion at site \( i \) in the spin state \( \sigma =\uparrow  \)
or \( \downarrow  \) and \( n_{i\sigma }=c^{\dagger }_{i\sigma }c_{i\sigma } \).
It is important to notice that the SL structure breaks particle-hole
symmetry.\cite{paivasantos4} The homogeneous
Hubbard model, in a grand-canonical ensemble description, is invariant
under a particle-hole transformation \( \left[ c_{i\sigma }\rightarrow
  c_{i\sigma }^{\dagger }(-1)^{i}\right]  \) 
only when \( \mu =U/2 \). In the superlattice case, a uniform chemical
potential cannot ensure this symmetry throughout the whole system.
Instead, under a particle-hole transformation the system is mapped
onto a different one with a spatially modulated chemical potential.

A weak coupling perturbation theory, similar to that for the homogeneous
model can be used to show that Eq.~(\ref{llsl}) indeed describes
the low energy and small momentum sector of the discrete model of
Eq.\ (\ref{hus}) in the limit of long layers; see the Appendix.
Then, in Eq.\ (\ref{llsl}) one has \( K_{\nu }(x)=K_{\lambda \nu } \)
and \( u_{\nu }(x)=u_{\lambda \nu } \) for \( x \) on the layer
\( \lambda =1,2 \), where \( K_{\lambda \nu } \) and \( u_{\lambda \nu } \)
are the usual uniform weak coupling LL parameters for each layer.
It is by now well established that a LL description is appropriate
for the low-energy sector of the Hubbard model, \emph{even in the
strong coupling limit} \( U\rightarrow \infty  \).\cite{schulz1}
Now, each long Hubbard sub-chain is still a finite-sized LL, though
connected to particle reservoirs at each end.\cite{castronetoetal} 
We therefore make the quite reasonable assumption
that the above LLSL description remains valid even in the strong coupling
limit. With respect to magnetic properties, the superlattice structure
(with repulsive interactions) does not break SU(2) symmetry, so that
the inhomogeneous \( K_{\sigma } \) is still expected to renormalize
to \( K_{\sigma }\rightarrow 1 \).

Because each sub-chain is an \emph{open} LL, there will be a certain
amount of charge redistribution between them, leading to a non-uniform
charge profile. Let us first consider the special case of two layers
only [with parameters \( \left( U_{1},t_{1}\right)  \) and \( \left(
U_{2},t_{2}\right)  \)]
initially disconnected and with the same initial density \( n=N/L \).
In general, these two sub-systems will not have the same chemical
potential. We then bring them in contact with each other, so that
particle exchange is allowed. Electrons will flow from one system
to the other until their chemical potentials exactly match:
\begin{equation}
\label{equi}
\mu(t_{1},U_{1},n_{1})=\mu(t_{2},U_{2},n_{2}),
\end{equation}
where \( \mu \) and \( n_{\lambda } \) are the chemical
potential and the equilibrium densities of each layer, respectively.
This is just the condition for thermodynamic equilibrium. Naturally,
conservation of total charge dictates that 
\begin{equation}
\label{cparnu}
n_{1}+\ell n_{2}=n(1+\ell ).
\end{equation}
 In order to determine \( n_{1} \) and \( n_{2} \), we must solve
simultaneously Eqs.~(\ref{equi}) and (\ref{cparnu}). The extension
to the case of more than two layers
leads to no modifications of
the above equations and the charge profile will be periodic with the
densities determined as above. 

The dependence of \( \mu \) on the density \( n \) and on the interaction
\( U \) can be obtained from the exact solution of the homogeneous
Hubbard model.\cite{liebwu} As a function of $n$, the chemical
potential $\mu(t,U,n)$ increases
monotonically and is discontinuous at half-filling, where it jumps from
$\mu_{-}(t,U)$ to $\mu_{+}(t,U)=U-\mu_{-}(t,U)$. Thus, the homogeneous
model is a Mott insulator at half-filling.  $\mu_{-}(t,U)$ is the lower
chemical potential at half-filling, given by\cite{liebwu}
\begin{equation}
\label{numen}
\mu _{-}(t,U)=2t-4t\int ^{\infty }_{0}\frac{J_{1}(\omega )d\omega }{\omega [1+e^{\frac{1}{2}\omega U/t}]},
\end{equation}
where $J_{1}(\omega )$ is a Bessel function.
To increase the particle number above half-filling, we need to pay an
energy given by
\begin{equation}
\Delta_H=\mu_+(t,U)-\mu_-(t,U)=U-2\mu_-(t,U),
\end{equation}
which is the quasiparticle gap. For later use, we also quote the
chemical potential of the non-interacting case,
\begin{equation}
\mu(t,0,n)=-2t\cos\left(\frac{\pi n}{2}\right).
\label{chem0}
\end{equation}

\begin{figure}
{\centering \resizebox*{3.in}{!}{\includegraphics*{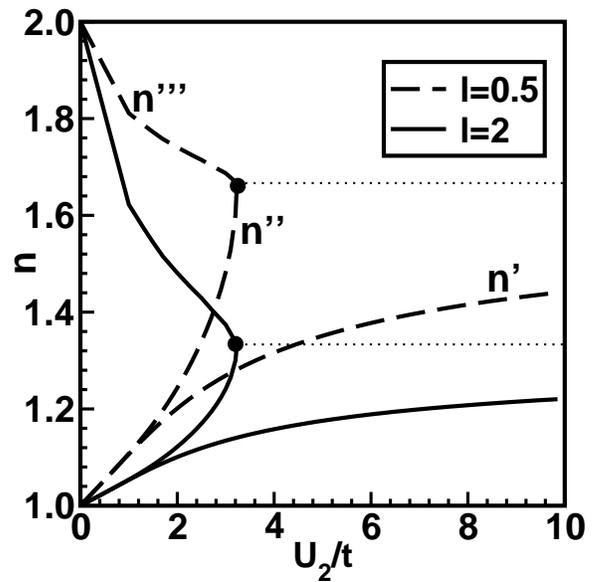}} \par}
\caption{
\label{hslfree}
Phase diagram of a Hubbard superlattice with \protect\( U_{1}=0\protect \)
and \protect\( t_{1}=t_{2}=t\protect \) for two layer length ratios:
\protect\( \ell =0.5\protect \) and \protect\( 2.\protect \) For each 
$\ell$, there are two
metallic phases and two insulating ones. 
The two dots locate $(U_c,n_c)$, where
\protect\(
U_{c}/t=3.2309\protect \) and \protect\( n_{c}=\left( 2+\ell \right)
/\left( 1+\ell \right) .\protect \)}
\end{figure} 

\subsection{The \protect\( U_{1}=0\protect \) case}

\label{u1=3D0}

We first consider the case in which one of the layers is `free' (\(
U_{1}=0 \)) and take \( t_{1}=t_{2}=t \) for simplicity. 
Figure \ref{hslfree} shows the phase diagram for \( \ell =L_{2}/L_{1}=0.5
\) and 2; the case $\ell =1$ has been discussed in Ref.\
\onlinecite{valencia1}. 
For the sake of comparison, one should also keep in mind the phase diagram
for the homogeneous LL, in which there is a single gapped (Mott)
insulating phase for any non-zero repulsion at half-filling; upon either
electron- or hole-doping the system becomes metallic.
In what follows, we start with a qualitative
discussion of the phase diagram, after which we provide the details of
how the boundaries and special points are determined.
\begin{figure}
{\centering \resizebox*{3.in}{!}{\includegraphics*{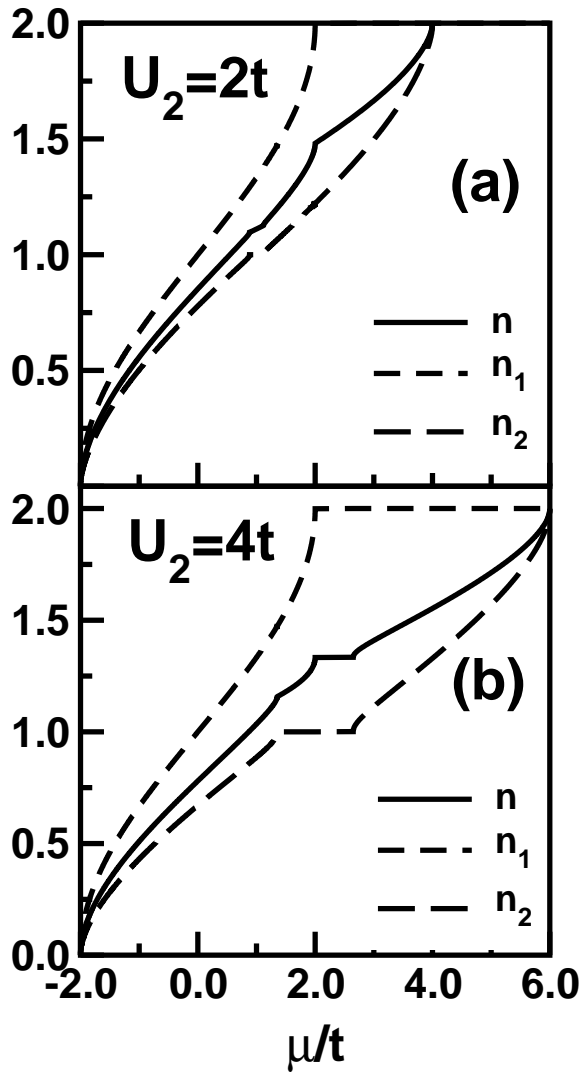}}
\par}
\caption{
\label{nXmu}
Particle densities for the $\ell=2$ Hubbard superlattice with $U_1=0$, as
functions of the chemical potential: $n$ is the overall charge density
(full lines), $n_1$ is the density at free sites (short-dashed curves),
and $n_2$ is the density at repulsive sites (long-dashed curves). Two
cases are considered: (a) $U_2=2t<U_c$, and (b) $U_2=4t>U_c$.}
\end{figure} 

In the case of a superlattice, while for $n<1$ the system is always
metallic, interesting metal-insulator
transitions have been found for $n>1$, as displayed in Fig.\
\ref{hslfree}.  Indeed, for a density $n$ just above half-filling, the
system is still metallic, with more particles occupying the free layer
than the repulsive one in order to decrease the overall electronic
repulsion: One has $n_1 > 1$ and $n_2 < 1$, as shown in Fig.\ \ref{nXmu}.
As the density is increased for given $\ell$ and $U_2$, electrons will be
accommodated in both layers without affecting the metallic character; see
Figs.\ \ref{hslfree} and \ref{nXmu}. This will persist until the repulsive
layer is half filled ($n_2=1$), when it becomes a \emph{Mott} insulator.
Recall that an insulating phase in one of the subsystems is signalled in
Fig.\ \ref{nXmu} by a horizontal plateau in the corresponding $n_i(\mu)$
($i=1,$ 2) plots. 
The system as a whole is therefore an \emph{insulator,} since it can be
thought of as a series arrangement of resistors.
However, the unusual fact is the \emph{gapless} nature of this insulating
phase: charge can be accommodated in the free layer at no energy cost,
since the system is compressible ($\partial n/\partial \mu \neq
0$) in this range of $n$; see Fig.\ \ref{nXmu}.

As the density is further increased, the system responds in two different
ways, depending on whether $U_2$ is larger or smaller than $U_c\equiv
3.2309t$ (for all $\ell$); see Figs.\ \ref{hslfree} and \ref{nXmu}. If
$U_2<U_c$ [Fig.\ \ref{nXmu}(a)], the insulating state can only be
sustained up to a limited amount of additional charge; that is, as long as 
it
is energetically favorable to accommodate this extra charge in the free
layer, while keeping $n_2=1$. Further increase in $n$ soon leads to an
increase in the occupation of the repulsive layer (with $2>n_1>n_2>1$) and
the system reenters an overall metallic phase. This metallic character
will be lost again for larger $n$, when the free layer becomes completely
full ($n_1=2, 1<n_2<2$), with the superlattice displaying
insulating behavior. Again, this insulating phase is gapless.

If $U_2>U_c$ [Fig.\ \ref{nXmu}(b)], on the other hand, all added electrons
will be accommodated in the free layer ($1<n_1<2, n_2=1$), so that the
superlattice remains in the state of a gapless insulator. Further increase
in the electron density leads to the free layer becoming a \emph{band}
insulator ($n_1=2$), while keeping the repulsive one pinned at
half-filling; the density, $n_c$, at which this occurs depends on the
aspect ratio, $\ell$, and is given by [\textit{c.f.,} Eq.~(\ref{cparnu})
and Refs.\ \onlinecite{paivasantos2, paivasantos4}] $n_c=\left( 2+\ell
\right)/\left(1+\ell \right)$.  Only at this special density does the
superlattice become a Mott insulator, since it is incompressible
($\partial n/\partial \mu = 0$); see Fig.\ \ref{nXmu}(b).  For $n>n_c$,
the free layer remains completely full, so that all added electrons go to
the repulsive layer; the superlattice behaves again as a gapless
insulator.

At this point it is worth commenting that the
phase diagram of Fig.\ \ref{hslfree} differs in two aspects from the one
found for \emph{thin} layers, obtained by means of Lanczos
diagonalizations: In Ref.\ \onlinecite{paivasantos2} no \emph{gapless}
insulating phases were probed, and the insulating phase for $n=n_c$ was
found to extend down to any $U_2>0$. 
The former difference is due to the fact that only gapped insulating
phases were probed, while the latter can be traced back to finite-size
effects. 

\begin{figure} 
{\centering
\resizebox*{3.in}{!}{\includegraphics*{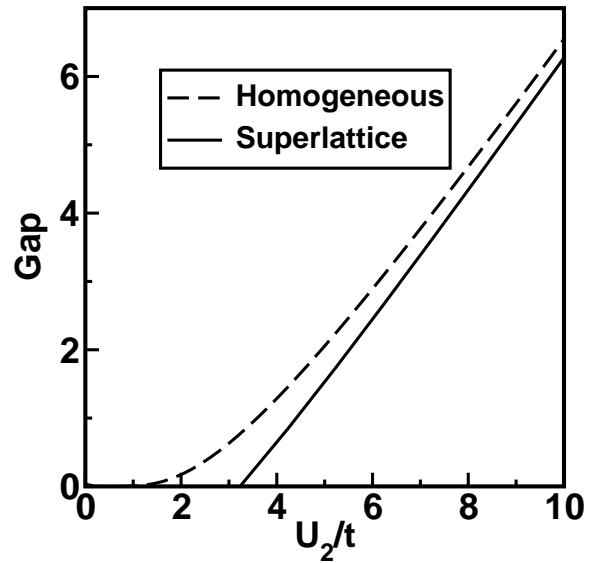}} \par}
\caption{\label{gap} The Mott-Hubbard gap at $n=n_c$ as a
function of the on-site coupling for both the homogeneous model 
(dashed line) and the superlattice (full line). This behavior 
is the same for all $\ell$.
}
\end{figure} 

Let us now fill in the details on how the lines and special points of
Fig.\ \ref{hslfree} are determined.
The dotted horizontal line at $n=n_c$ is obtained by setting 
both \( n_{1}=2 \) and \( n_{2}=1 \).
However, this condition can only be obtained if \( U_{2}>U_{c}, \)
where \( U_{c} \) is determined implicitly by (see Fig.\ \ref{nXmu})
\begin{equation}
\label{uc}
\mu_+(t,U_c)=\mu(t,0,n_1=2)=2t. 
\end{equation}
Since neither Eq.\ (\ref{uc}) nor Eq.\ (\ref{numen}) depends on $\ell$,
this condition yields the same \( U_{c}\approx 3.23097t \) for any
finite aspect ratio.   

Besides, for \( U_{2}<U_{c} \), the system is always gapless. For \(
U_{2}>U_{c} \) and \( n=n_{c}, \) the system shows a Mott-Hubbard gap
given by the energy difference between the highest occupied state,
which is the upper edge of the non-interacting band at $2t$, and the
lowest unoccupied level, which is the higher chemical potential of the
half-filled Hubbard chain at $\mu_+(t,U_2)$
\begin{equation}
\label{gap1}
\Delta _{S}=\mu_+(t,U_2)-2t=U_{2}-2t-\mu _{-}(t,U_{2}).
\end{equation}
For the one-dimensional Hubbard model, one has \( \Delta _{H}\sim
(8\sqrt{tU}/\pi )\exp (-2\pi t/U) \) in weak coupling and \( \Delta
_{H}\propto U \) in strong coupling. For the HSL, we found that \( \Delta
_{S} \) is linear with \( U_{2} \) for large \( U_{2} \) and is always
lower than the gap of the corresponding homogeneous system; see Fig.~\ref{gap}.

\begin{figure} 
{\centering
\resizebox*{3.in}{!}{\includegraphics*{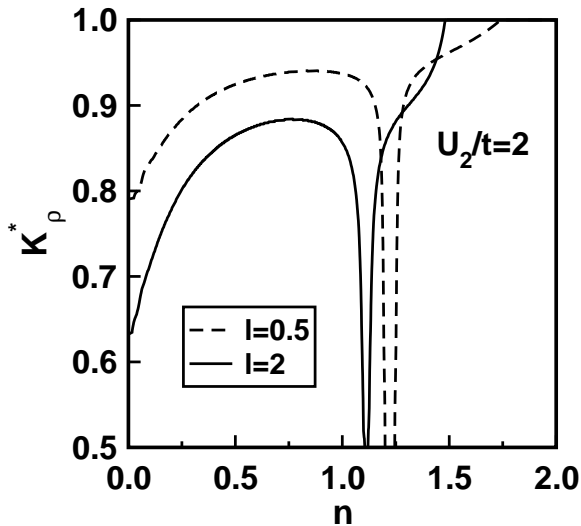}} \par}
\caption{\label{krhofree}The correlation exponent \protect\( K_{\rho
}^{\ast }\protect \) as a function of the density \protect\( n\protect \)
for $U_1=0$, \protect\( U_{2}/t=2\protect \) and \protect\( \ell
=0.5,2\protect \).
}
\end{figure} 

The two metallic phases are characterized by \( n_{1}<2,n_{2}<1 \)
(lower one) and \( n_{1}<2,n_{1}>1 \) (upper one). The metal-insulator
transition (MIT) lines can therefore be obtained by means of
Eqs.~(\ref{equi}), (\ref{cparnu}), the Lieb-Wu chemical potential \(
\mu \left(t,U_{2},n_{2}\right)\) and (\ref{chem0}).  Therefore, in
Fig.~\ref{hslfree}, (i) \( n' \) is the line in which the lower
Hubbard band of the interacting sub-chain becomes fully occupied, (ii)
\( n''\) is the one in which the upper Hubbard band starts to fill,
and (iii) \( n'''\) is the line in which the non-interacting sub-chain
fills up. Thus,
\begin{eqnarray}
\mu _{-}\left( t,U_{2}\right)  & = & -2t\cos \left( \frac{\pi }{2}\left[ \left( 1+\ell \right) n-\ell \right] \right) ,\label{line1} \\
\mu _{+}\left( t,U_{2}\right)  & = & -2t\cos \left( \frac{\pi }{2}\left[ \left( 1+\ell \right) n-\ell \right] \right) ,\label{line2} \\
2t & = & \mu\left( t,U_{2},\frac{\left( 1+\ell \right) n-2}{\ell }\right).\label{line3} 
\end{eqnarray}

The LL description of Sec.\ \ref{model} is only valid in the metallic
regions of the phase diagram, where no gap is present in either the
spin or the charge sectors. In these regions, we have
\begin{eqnarray*}
c_{\rho } & = & \frac{v_{F}\left( 1+\ell \right) }{\sqrt{1+\Delta _{\rho }\ell v_{F}/u_{2,\rho }+\left( \ell v_{F}/u_{2,\rho }\right) ^{2}}},\\
c_{\sigma } & = & \frac{v_{F}\left( 1+\ell \right) }{1+\ell v_{F}/u_{2,\rho }},
\end{eqnarray*}
where \( v_{F} \) is the Fermi velocity. When the insulating phase is
approached from the lower metallic region (see Fig.~\ref{hslfree}), \(
c_{\rho }\rightarrow 0 \) as a result of \( u_{2,\rho }\rightarrow 0 \) in
the interacting layer. In Fig.\ \ref{krhofree}, we show the effective
exponent $K_\rho^\ast$ as a function of the filling \( n \). For any \(
\ell \), both metallic phases have \( 1/2<K_{2,\rho }<K_{\rho }^{\ast }<1
\) and the charge and spin correlation functions decay faster than in the
homogeneous system. This is a direct consequence of the `weighted average'
character of the effective exponent \( K_{\rho }^{\ast } \).  By the same
token, for a given $n$ on the lower metallic phase, $K_\rho^\ast$
decreases as \( \ell \) increases.  In the upper metallic phase,
$K_\rho^\ast$ always tends to the non-interacting value of 1 as the upper
insulating region is approached; for the superlattice with larger $\ell,$
$K_\rho^\ast$ reaches 1 at a lower overall density.

\begin{figure}
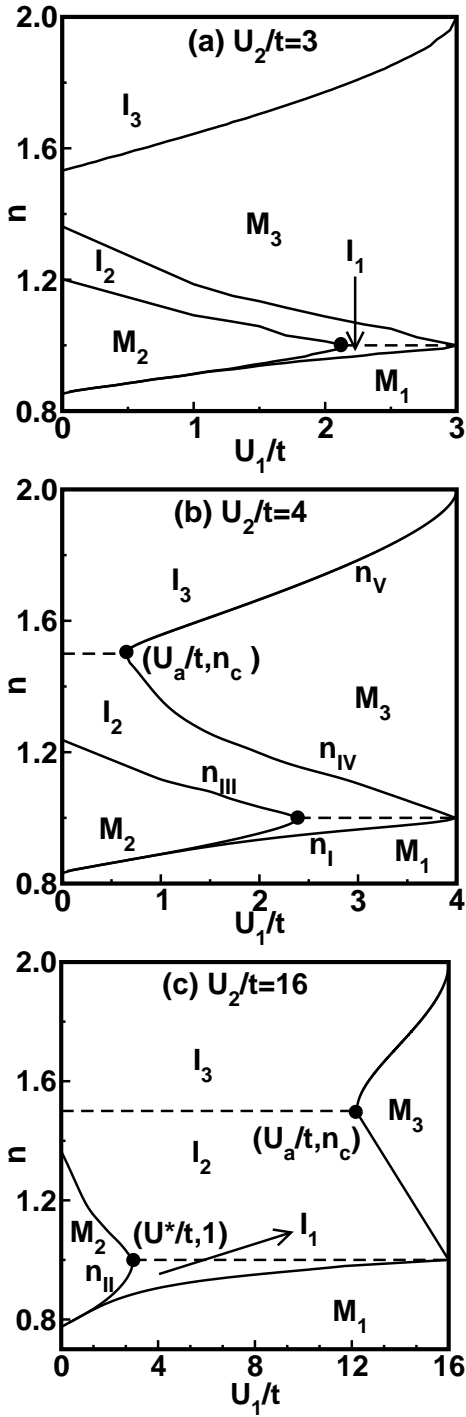

{\centering \resizebox*{2.4in}{!}{\includegraphics*{fig07a.eps}} \par}
{\centering \resizebox*{2.4in}{!}{\includegraphics*{fig07b.eps}} \par}
{\centering \resizebox*{2.4in}{!}{\includegraphics*{fig07c.eps}} \par}
\caption{\label{tph16}
Phase diagram for a $\ell=1$ ($n_c=1.5$) Hubbard superlattice with
$U_1>0$, and three different values of $U_2$. In each case there are three
metallic (M) phases, and three \emph{gapless} insulating (I) phases;
\emph{gapped} insulating phases appear as horizontal dashed lines.
In (a), \protect\(U_{2}=3t\protect\) and $U^*\approx 2.15t$ (see text);
in (b), \protect\(U_{2}=4t\protect\), $U^*\approx 2.39t$, and $U_a\approx
0.64t$; and in (c), \protect\(U_{2}=16t\protect \), $U^*\approx 3.02t$,
and $U_a\approx 12.2t$.}
\end{figure} 

\subsection{The general case: \protect\( U_{2}\geq U_{1}\neq 0\protect \)}

\label{hslnotfree}

We now consider a more general HSL, with different non-vanishing
coupling constants on each layer (\( U_{1}\neq U_{2} \)), while
keeping the same hopping amplitude \( t \) throughout the lattice
(Fig.~\ref{slscheme}(a)). Using once again the exact expression for
the chemical potential as a function of both \( U \) and \( n
\),\cite{liebwu} we have determined the charge profile of the
superlattice system. The charge tends to accumulate in the layer with
the smaller coupling, which we choose to call layer 1. This is rather
intuitive, since electrons decrease their mutual repulsion energy by
flowing into the less interacting layer.

The phase diagram for this HSL is very rich.  We observe six different
phases, three metallic (M$_1$, M$_2$ and M$_3$) and three insulating
(I$_1$, I$_2$ and I$_3$), each characterized by its charge profile, as
shown for three illustrative cases in Fig.~\ref{tph16}. The topology
of the phase diagram is the same for any \( \ell \) and the limiting
cases \( U_{1}\rightarrow 0 \) (Section \ref{u1=3D0}) and \(
U_{1}=U_{2} \) (homogeneous chain) are recovered. On each phase
diagram of Fig.\ \ref{tph16}, there are five MIT lines, labeled by
$n_I$ through $n_V$, which are determined similarly to the case
$U_1=0$ discussed before (see Table~\ref{table1}). We get:

\noindent\textit{line I}: 
\begin{equation}
\label{line1b}
\mu _{-}(t,U_{1})=\mu \left(t,U_{2},\frac{(1+\ell )n-1}{\ell
}\right);
\end{equation}
\textit{line II}: 
\begin{equation}
\label{line2b}
\mu _{+}(t,U_{1})=\mu \left(t,U_{2},\frac{(1+\ell )n-1}{\ell
}\right);
\end{equation}
\textit{line III}: 
\begin{equation}
\label{line3b}
\mu \big(t,U_{1},(1+\ell )n-\ell \big)=\mu _{-}(t,U_{2});
\end{equation}
\textit{line IV}: 
\begin{equation}
\label{line4b}
\mu \big(t,U_{1},(1+\ell )n-\ell \big)=\mu _{+}(t,U_{2});
\end{equation}
\textit{line V}: 
\begin{equation}
\label{line5b}
\mu (t,U_{1},n_{1}=2)=\mu _{-}\left(t,U_{2},\frac{(1+\ell )n-2}{\ell
}\right).
\end{equation}
For \( U_{1}=0 \), the lines \( n_{III},\ n_{IV}, \) and \( n_{V} \)
determine the phase diagram of Section \ref{u1=3D0} 
[Eqs.~(\ref{line1})--(\ref{line3})].\cite{valencia1}

One of the consequences of a non-zero $U_1$ is to push the lower
metallic phase of Fig.\ \ref{hslfree} to smaller densities, as shown in
Fig.~\ref{tph16} (M$_1$). In addition to this phase, which spans all
values of $U_1<U_2$, there are two other metallic regions (M$_2$ and
M$_3$). And in-between metallic phases, one finds insulating phases,
one of which (I$_1$) is now stable for $n<1$, unlike the case for
$U_1=0$.  These insulating phases have either \( n_{\lambda }=1,\ 
\lambda =1,2, \) or \( n_{1}=2 \) (see Table~\ref{table1}). Once
again, there is a `division of labor' between the two 
types of sub-chains: while one is gapped (Mott) or completely filled
(band), being responsible for the insulating behavior of the system,
the other remains gapless and so does the system as a whole.

\begin{table}
\caption{\label{table1} This table lists the various metallic
(M$_1$, M$_2$ and M$_3$) and insulating phases (I$_1$, I$_2$ and
I$_3$) of Fig.~\ref{tph16} with the corresponding sub-chain
densities. The last column shows the nature of the transition lines
between the phases ($n_I$ through $n_V$ in Fig.~\ref{tph16}). 
LHB $\lambda$ and UHB $\lambda$ respectively stand for 
lower Hubbard band and upper Hubbard band in layer $\lambda=1,2$.
}
\begin{ruledtabular}
\begin{tabular}{ccc}
& Sub-chain densities & Transition line \\
\hline
M$_1$ & $n_1<1$, $n_2<1$ & - \\ \hline
$\Downarrow$ & - & LHB 1 fills up ($n_{I}$) \\ \hline
I$_1$ & $n_1=1$, $n_2<1$ & - \\ \hline
$\Downarrow$ & - & UHB 1 starts to fill ($n_{II}$)\\ \hline
M$_2$ & $n_1>1$, $n_2<1$ & - \\ \hline
$\Downarrow$ & - & LHB 2 fills up ($n_{III}$)\\ \hline
I$_2$ & $n_1>1$, $n_2=1$ & - \\ \hline
$\Downarrow$ & - & UHB 2 starts to fill ($n_{IV}$)\\ \hline
M$_3$ & $n_1>1$, $n_2>1$ & - \\ \hline
$\Downarrow$ & - & UHB 1 fills up ($n_{V}$)\\ 
\end{tabular}
\end{ruledtabular}
\end{table}

Figure \ref{tph16}(a) shows the phase diagram for \( U_{2}=3t<U_{c} \)
($U_c$ is the same as for the case $U_1=0$).  The HSL has a gap at the
density \( n=1 \) for \( U_{1}>U^{*}\approx 2.145608t \); this $n=1$ line
separates the I$_1$ (i.e., \( n_{1}=1 \), \( n_{2}<1 \)) and the I$_2$
(\( n_{1}>1 \), \( n_{2}=1 \)) gapless insulating phases. For \(
U_{1}<U^{*} \) one goes through a sequence of MIT's, in which all
insulating phases are gapless.

In Fig.~\ref{tph16}(b), we show the phase diagram for \(
U_{2}=4t>U_{c} \). As the overall density is increased from 1 in the
interval \( U_{a}<U_{1}<U^{*} \), where \( U_{a}\approx 0.6433t \) and
\( U^{*}\approx 2.39149t \), the system goes through a sequence of
MIT's without ever being gapped.  However, for \(U_{1}<U_{a} \), the
intermediate I$_2$ (\( n_{1}>1 \), \( n_{2}=1 \)) and the I$_3$ (\(
n_{1}=2 \), \( n_{2}>1 \)) gapless insulating phases are separated by
the dashed line at the density \( n_{c}=(2+\ell )/(1+\ell ) \), where
the system is fully gapped. Similarly, another gap appears at the
density \( n=1 \) for \( U_{1}>U^{*} \), which again separates gapless
insulating phases I$_1$ (\( n_{1}=1 \), \( n_{2}<1 \)) and I$_2$ (\(
n_{1}>1 \), \( n_{2}=1 \)).

For \( U_{2}=16t>U_{c} \) [Fig.~\ref{tph16}(c)] and \( U^{*}<U_{1}<U_{a}
\) (now \( U^{*}=3.01509t \) and \( U_{a}=12.1724t \)) the system is
metallic only below line $I$, which approaches $n=1$ for large $U_1$;
also, gapped behavior is again observed at densities \( n=1 \) and \(
n_{c}\), with all other insulating phases being
gapless. For each of the regions \( U_{1}<U^{*} \) and \( U_{1}>U_{a} \),
a `tipped' metallic phase is observed.

\begin{figure}
{\centering \resizebox*{3.in}{!}{\includegraphics*{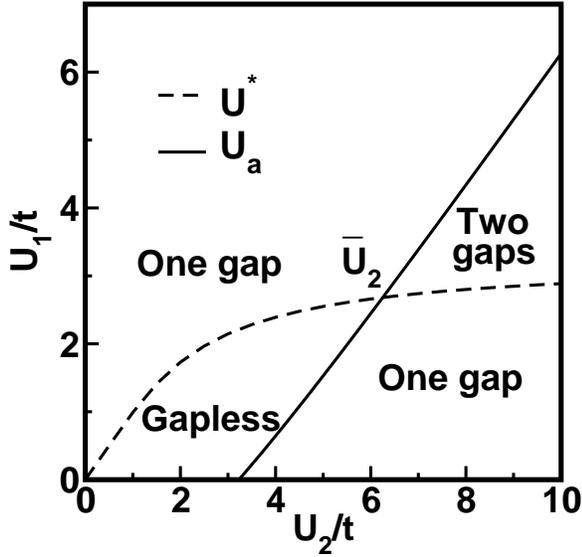}} \par}
\caption{\label{tUs}Parameters \protect\( U^{*}\protect \) and \protect\( U_{a}\protect \)
as functions of \protect\( U_{2}\protect \). The two curves meet
at \protect\( \overline{U}_{2}=6.25261t.\protect \)}
\end{figure} 

The above discussion indicates that there are special values of $U_1$, \(
U^{*} \) and \( U_{a} \), which respectively represent the `tip' positions
of the low- and high-density metallic phases.
Their dependence on \( U_{2} \) can be extracted from the solutions of 
\begin{equation}
\mu_+(t,U_2)=\mu(t,U_a,n_1=2)=2t+U_a 
\end{equation}
and of
\begin{equation}
\mu_+(t,U^*)=\mu _{+}(t,U_{2}),
\end{equation}
and are shown in Fig.~\ref{tUs}.
It should be noted that these values are independent of the aspect ratio
\( \ell \).  As Fig.~\ref{tUs} reveals, one should not be misled by the
different horizontal scales in Fig.~\ref{tph16}: the low-density tip does
not recede as $U_2$ increases, since \( U^{*} \) actually increases
monotonically with $U_2$, saturating at \( U_{c} \) as \(U_{2}\to \infty
\). On the other hand, Fig.~\ref{tUs} shows that \( U_{a} \) is only
defined above a certain threshold, $U_2=U_c$, reflecting the fact that
when the coupling in layer 2 is small, the situation $n_1=2,\ n_2=1$
is never realized; above $U_c$, $U_a$ increases linearly with \( U_{2}\).

According to our previous analyses, these two curves
(which intersect at \( \overline{U}_{2}=6.25261t \)) define regions in the
\( (U_{1},U_{2}) \) plane characterized by the number of gaps in
the sub-units for appropriate fillings, as specified in Fig.~\ref{tUs}.

\begin{figure}
{\centering \resizebox*{3.in}{!}{\includegraphics*{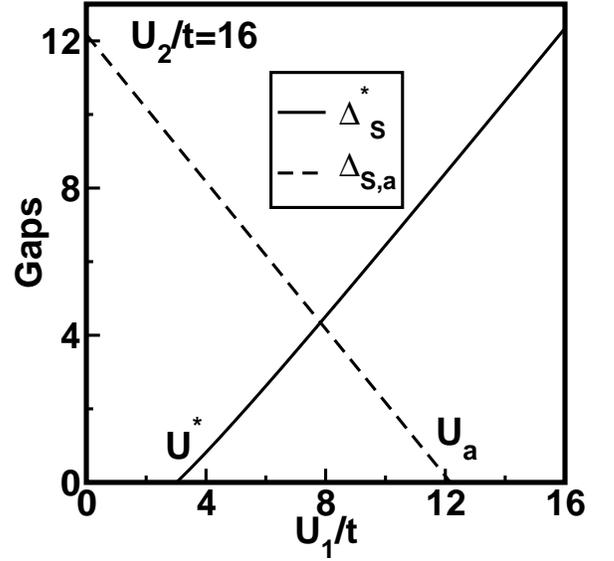}} \par}
\caption{\label{gaps16}The Mott-Hubbard gaps \protect\( \Delta _{S}^{*}\protect \)
and \protect\( \Delta _{S,a}\protect \) at the densities \protect\( n=1\protect \)
and \protect\( n_{c}=\left( 2+\ell \right) /\left( 1+\ell \right) \protect \),
respectively. Here, \protect\( U_{2}=16t\protect \) and \protect\( \ell =1\protect \).}
\end{figure}

Similarly to the case $U_1=0$, the gaps at the densities \( n=1 \) and \(
n=n_{c}\) are given, respectively, by 
\begin{eqnarray} \Delta _{S}^{*}
& = & \mu _{+}(t,U_{1})-\mu _{-}(t,U_{2}), \label{gap2} \\
\Delta _{S,a} & = & \mu _{+}(t,U_{2})-2t-U_{1}, 
\end{eqnarray} 
and, again, they do not depend on \( \ell \). The gaps \( \Delta _{S}^{*}
\) and \( \Delta _{S,a} \), for \( U_{2}=16t \), are shown in
Fig.~\ref{gaps16} as functions of \( U_{1} \). The gap at \( n=1 \) [\(
n=n_{c} \)] increases [decreases] linearly with \( U_{1}
\) and vanishes for \( U_{1}<U^{*} \) [\( U_{1}>U_{a} \)].

\begin{figure}
{\centering \resizebox*{3.in}{!}{\includegraphics*{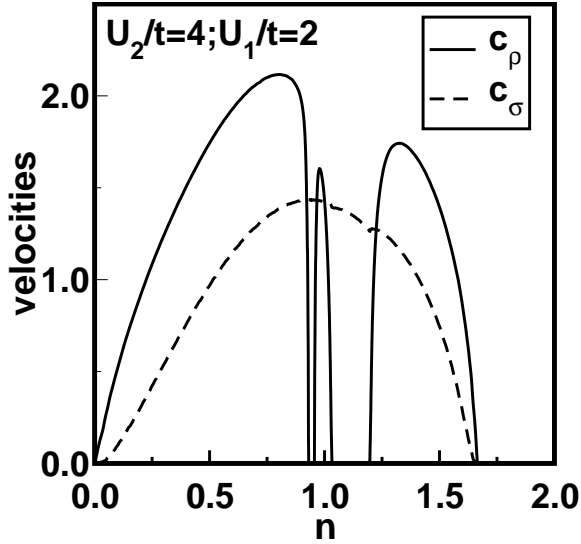}} \par}
\caption{\label{vel}The effective charge and spin velocities \protect\( c_{\rho }\protect \)
(full line) and \protect\( c_{\sigma }\protect \) (dashed line) for
a Hubbard superlattice as a function of \protect\( n\protect \) for
\protect\( U_{1}=2t\protect \), \protect\( U_{2}=4t\protect \) and
\protect\( \ell =1\protect \).}
\end{figure} 

For the Hubbard model with repulsive interactions we have \( u_{\sigma
}\leq v_{F} \) and \( u_{\rho }>v_{F} \).\cite{schulz1} For \( U_{2}=4t \)
and \( U_{1}=2t \), the effective charge and spin velocities for the \(
\ell =1 \) one-dimensional Hubbard superlattice are shown in
Fig.~\ref{vel} as functions of \( n \). The effective charge velocity
(full line in Fig. \ref{vel}) vanishes upon approaching the insulating
regions as a result of the vanishing charge velocities of the individual
sub-chains \( u_{\lambda ,\rho }\rightarrow 0 \). Thus, \( c_{\rho } \)
shows a re-entrant behavior as a function of \( n \) (cf.
Fig.~\ref{tph16}). As in the homogeneous case, the effective spin velocity
is always smaller than the Fermi velocity and only vanishes in the upper
insulating phase (dashed line in Fig.\ \ref{vel}). The different
behaviors of \( c_{\rho } \) and \( c_{\sigma } \) can be traced back to
the the fact that $K_\rho^\ast$ is sensitive to the superlattice
structure, while \( K_{\sigma }^\ast=1 \), since
$K_{1\sigma}=K_{2\sigma}=1$ as a result of the SU(2) symmetry being
preserved.

The preservation of SU(2) symmetry also leads to \( \overline{K}_{\sigma
}=K^{*}_{\sigma }=1 \). Thus, from Eqs.~(\ref{dens-dens-corr}) and
(\ref{spin-spin-corr}), the density-density and spin-spin correlation
functions for the HSL are dominated by \( \langle O^{\dagger }O\rangle\sim
\left| x-y\right| ^{-1-K_{\rho }^{\ast }} \). These terms correspond to \(
2k_{F} \)-CDW and \( 2k_{F} \)-SDW in the homogeneous system. Here, \(
K_{2\rho }<K_{\rho }^{\ast }<K_{1\rho } \) and the density-density and
spin-spin correlation functions for the HSL decay faster (slower) than for
a homogeneous system with \( U=U_{2} \) (\( U=U_{1} \)). Similarly,
pairing correlation functions are \( \langle O^{\dagger }O\rangle\sim
\left| x-y\right| ^{-1-\overline{K}_{\rho }} \). In spite of the presence
of effective exponents \( K_{\rho }^{\ast } \) and \( \overline{K}_{\rho }
\), the condition for superconducting quasi-long range order is again
\( K_{\rho }^{\ast } > 1\), analogous to the homogeneous case;
this condition, nonetheless, remains unsatisfied.

\begin{figure}
{\centering \resizebox*{3.in}{!}{\includegraphics*{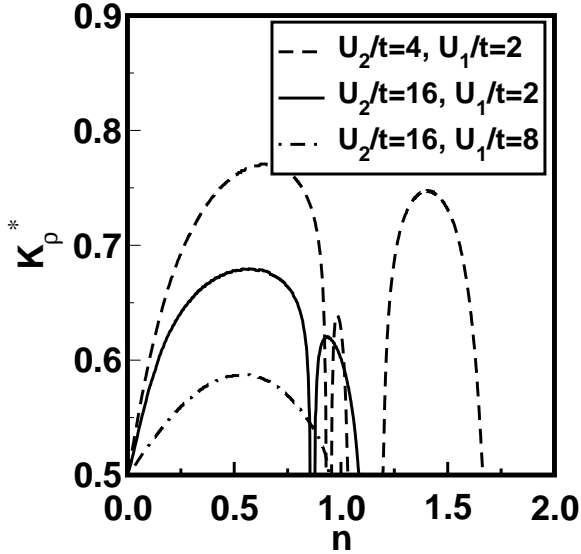}} \par}
\caption{\label{Krhon}The correlation exponent \protect\( \mathrm{K}_{\rho }^{\ast }\protect \)
as a function of the band filling \protect\( n\protect \) for \protect\( \ell =1\protect \)
and several values of the coupling constants. }
\end{figure} 

In Fig.~\ref{Krhon}, the correlation exponent \( K_{\rho }^{\ast } \)
of the HSL is shown as a function of band filling, for different \( \ell
=1 \) superlattices: HSL-1 with \( U_{2}=4 \)t and \( U_{1}=2t \);
HSL-2 with \( U_{2}=16t \) and \( U_{1}=2t \); HSL-3 with \( U_{2}=16t \)
and \( U_{1}=8t \). For any \( \ell  \), all metallic phases are
characterized by \( 1/2<K_{\rho }^{\ast }<1 \). We note that
HSL-1 has three metallic phases, HSL-2 has two metallic phases
and HSL-3 has only one metallic phase. On the low density side,
\( K_{\rho }^{\ast } \) approaches 1/2 in contrast to the case \( U_{1}=0 \)
(Sec.~\ref{u1=3D0}), in which \( K_{\rho }^{\ast } \) remains between
1/2 and 1. From Eq.\ (\ref{knustar}) one sees that \( K_{\rho }^{\ast } \)
interpolates monotonically between \( K_{1\rho } \) and \( K_{2\rho } \)
as \( \ell  \) is varied from 0 to \( \infty  \), highlighting the
possibility of a continuous `modulation' of a physical parameter
through the tuning of the superlattice structure.

\subsection{Two different hoppings: \protect\( t_{2}\geq t_{1}\protect \) and
\protect\( U_{\lambda }=U>0\protect \)}

\label{hslt1t2}

We now consider two Hubbard chains arranged periodically with the same
coupling \( U_{\lambda }=U>0 \), but different hoppings \( t_{2}>t_{1}
\) (Fig.~\ref{slscheme}(b)).\cite{cabraetal2}

\begin{figure}
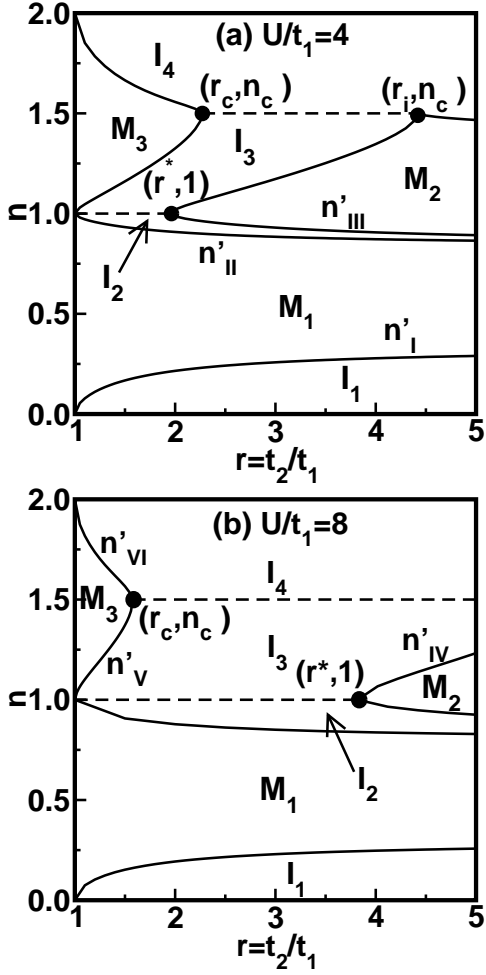

{\centering \resizebox*{2.5in}{!}{\includegraphics*{fig12a.eps}} \par}
{\centering \resizebox*{2.5in}{!}{\includegraphics*{fig12b.eps}} \par}
\caption{\label{pht}Phase diagram for a Hubbard superlattice with different
hoppings \protect\( t_{1}\protect \) and \protect\( t_{2}\protect \), in 
terms of filling \protect\( n\protect \) and hopping ratio $t_2/t_1$;
the on-site coupling is homogeneous and set to \protect\( 
U=4t_{1}\protect \) in 
(a) and \protect\( U=8t_{1}\protect \) in (b). There are 
three metallic (M) and four insulating
(I) phases (\protect\( \ell =1\protect \) and \protect\( n_{c}=(2+\ell )/(1+\ell )\protect \)
). In (a), \protect\( r^{*}=1.94847\protect \), \protect\( r_{c}=2.26984\protect \)
and \protect\( r_{i}=4.4227\protect \) and in (b), \protect\( r^{*}=3.818576\protect \),
\protect\( r_{c}=1.57735\protect \) and \protect\(
r_{i}=6.02322\protect \). Besides, the values of $n^\ast$ are (see
text for definition) (a) $n^\ast = 0.670511$ and (b) $n^\ast =
0.594067$.
}
\end{figure}

Initially, the charge tends to accumulate in the layer with larger
hopping (layer 2), because its chemical potential is the smallest.
Eventually, their chemical potentials become equal at the special
density $n^\ast$, determined by $\mu (t_\lambda,U,n^\ast)=0$.  Then,
for \( n>n^{\ast } \), the charge flow is reversed and proceeds from
layer 2 to layer 1. \( n^{\ast } \) is independent of \( r=t_{2}/t_{1}
\) and \( \ell \), and decreases with \( U \) (see Fig.~\ref{pht}).

It is interesting to plot a phase diagram in terms of the density and
the ratio between the two hopping amplitudes, $r\equiv t_2/t_1$. We
then identify seven different phases, three metallic (M$_1$, M$_2$ and
M$_3$) and four insulating (I$_1$, I$_2$, I$_3$ and I$_4$), as shown
in Fig.~\ref{pht} for \( U=4t_{1} \) and \( U=8t_{1} \) with \( \ell
=1 \) and listed in Table~\ref{table2}. The value of $n^\ast$ lies
within the M$_1$ phase. We mention that several different insulating
and metallic phases were also found in a p-merized Hubbard chain in a
magnetic field.\cite{cabraetal2}

\begin{table}
\caption{\label{table2} This table lists the various metallic
(M$_1$, M$_2$ and M$_3$) and insulating phases (I$_1$, I$_2$,
I$_3$ and I$_4$) of Fig.~\ref{pht} with the corresponding sub-chain
densities. The last column shows the nature of the transition lines
between the phases ($n'_I$ through $n'_{VI}$ in Fig.~\ref{pht}). 
LHB $\lambda$ and UHB $\lambda$ respectively stand for 
lower Hubbard band and upper Hubbard band in layer $\lambda=1,2$.
}
\begin{ruledtabular}
\begin{tabular}{ccc}
& Sub-chain densities & Transition line \\ \hline
I$_1$ & $n_1=0$, $n_2<1$ & - \\ \hline
$\Downarrow$ & - & LHB 1 starts to fill ($n'_{I}$) \\ \hline
M$_1$ & $n_1<1$, $n_2<1$ & - \\ \hline
$\Downarrow$ & - & LHB 1 fills up ($n'_{II}$) \\ \hline
I$_2$ & $n_1=1$, $n_2<1$ & - \\ \hline
$\Downarrow$ & - & UHB 1 starts to fill ($n'_{III}$)\\ \hline
M$_2$ & $n_1>1$, $n_2<1$ & - \\ \hline
$\Downarrow$ & - & LHB 2 fills up ($n'_{IV}$)\\ \hline
I$_3$ & $n_1>1$, $n_2=1$ & - \\ \hline
$\Downarrow$ & - & UHB 2 starts to fill ($n'_{V}$)\\ \hline
M$_3$ & $n_1>1$, $n_2>1$ & - \\ \hline
$\Downarrow$ & - & UHB 1 fills up ($n'_{VI}$)\\ \hline
I$_4$ & $n_1=2$, $n_2>1$ & - \\
\end{tabular}
\end{ruledtabular}
\end{table}

Following the same reasonings as before, the lines in the phase diagram in
Fig.~\ref{pht} are given by:

\noindent
\textit{line I}:
\begin{equation}
\mu(t_{1},U,n_{1}=0)=\mu\left(t_{2},U,\frac{n'_{I}(1+\ell )}{\ell}\right),
\label{Ip}
\end{equation}
\textit{line II}:
\begin{equation}
\mu _{-}(t_{1},U)=\mu \left(t_{2},U,\frac{n'_{II}(1+\ell )-1}{\ell} 
\right),
\label{IIp}
\end{equation}
\textit{line III}: \begin{equation}
\mu _{+}(t_{1},U)=\mu \left(t_{2},U,\frac{n'_{III}(1+\ell )-1}{\ell} 
\right),
\label{IIIp}
\end{equation}
 \textit{line IV}: \begin{equation}
\mu \left(t_{1},U,n'_{IV}(1+\ell )-\ell \right)=\mu_{-}(t_{2},U),
\label{IVp}
\end{equation}
 \textit{line V}: \begin{equation}
\mu \left(t_{1},U,n'_{V}(1+\ell )-\ell \right)=\mu_{+}(t_{2},U),
\label{Vp}
\end{equation}
 \textit{line VI}: \begin{equation}
\mu(t_{1},U,n_{1}=2)=\mu \left(t_{2},U,\frac{n'_{VI}(1+\ell )-2}{\ell} 
\right).
\label{VIp}
\end{equation}
\noindent
Again, the topology of the phase diagrams in Fig.~\ref{pht} is the
same for any \( \ell  \).

At small densities (I$_1$ phase), charge accumulates in layer 2 while
layer 1 is empty ($n_1=0$); the system is therefore a gapless
insulator.

\begin{figure}
{\centering \resizebox*{3.in}{!}{\includegraphics*{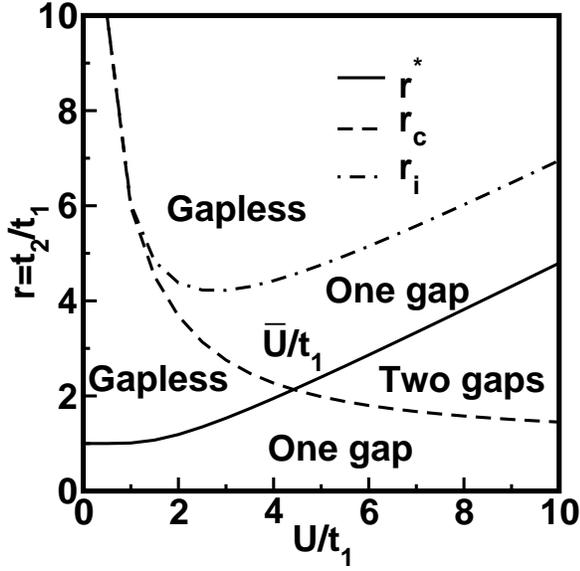}} \par}
\caption{\label{tsu}Parameters \protect\( r^{*},\protect \) \protect\(
r_{c}\protect \) and \protect\( r_{i}\protect \) as functions of
\protect\( U/t_{1}\protect \). Here, \protect\(
\overline{U}=4.4191t_{1}\protect \).} 
\end{figure}

\begin{figure}
{\centering \resizebox*{3.in}{!}{\includegraphics*{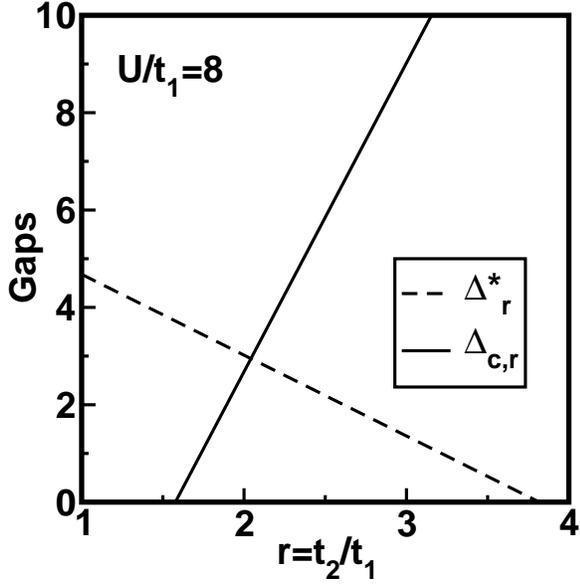}} \par}
\caption{\label{tgap}The Mott-Hubbard gaps \protect\( \Delta 
_{r}^{*}\protect \) and
\protect\( \Delta _{c,r}\protect \) at densities \protect\( n=1\protect \)
and \protect\( n_{c}=(2+\ell )/(1+\ell )\protect \), respectively.
Here, \protect\( U=8t_{1}\protect \) and \protect\( \ell =1\protect \).}
\end{figure}

As the density increases, layer 1 only starts being filled at 
$n=n_I^\prime
(r)$, determined by Eq.\ (\ref{Ip}), which locates a transition to a
metallic state (M$_1$); see Fig.\ \ref{pht}. Further increase in the overall
density leads to an increase in both $n_1$ and $n_2$. When layer 1 becomes
half-filled, which occurs at $n=n_{II}^\prime(r)$ as determined from Eq.\
(\ref{IIp}), the system reenters a gapless insulating state (I$_2$). If $r<r^*$,
where
\begin{equation}
r^{*} = \frac{\mu_{+}(t_{1},U)}{\mu_{-}(t_{2},U)},
\label{rstar}
\end{equation}
upon increasing the density the system goes through a gapped phase at $n=1$. 
The dependences of $r^*$ with $U$, and of the gap at $n=1$,
\begin{equation}
\Delta_{r}^{*} = \mu_{+}(t_{1},U)-\mu_{-}(t_{2},U),
\label{deltastar}
\end{equation}
with $r$, are shown in Figs.\
\ref{tsu} and \ref{tgap}, respectively; note that \( r^{*}(U=4t_1)=1.94847
\) and \( r^{*}(U=8t_1)=3.818576 \). By contrast, if $r>r^*$, the system
enters a metallic phase (M$_2$) bounded by $n_{III}^\prime(r)$, and
$n_{IV}^\prime$, given by Eqs.\ (\ref{IIIp}) and (\ref{IVp}).

When increasing the density above half filling, the sequence of
phases depends crucially on whether $r^*$ is smaller or larger than 
\begin{equation}
r_{c} = \frac{\mu(t_{1},U,n_{1}=2)}{\mu_{+}(t_{2},U)},
\label{rc}
\end{equation}
which, according to Fig.\ \ref{tsu}, occurs when $U<\overline{U}\approx
4.4191t_{1}$ or when $U>\overline{U},$ respectively. 

Let us first consider $U<\overline{U}$, which is the situation of Fig.\
\ref{pht}(a). If $r<r^*<r_c$, one goes through two transitions as $n$
increases: I$_3 \rightarrow$ M$_3$ at $n_{V}^\prime$ [see Eq.\ (\ref{Vp})],
and M$_3 \rightarrow$ I$_4$ at $n_{VI}^\prime$ [Eq.\ (\ref{VIp})]. If
$r^*<r<r_c$, the sequence is M$_2$-I$_3$-M$_3$-I$_4$, until the
lattice is completely filled.  Another regime is determined by
\begin{equation}
r_{i} = \frac{\mu (t_{1},U,n_{1}=2)}{\mu _{-}(t_{2},U)},
\label{ri}
\end{equation}
whose dependence on $U/t_1$ is also shown in Fig.\ \ref{tsu}.
If $r_c< r < r_i$, the system goes from a metallic (M$_2$) to a
gapless insulating phase (I$_3$), and then, at $n=n_c$, another
Mott-Hubbard gap opens, which is given by 
\begin{equation}
\Delta_{c,r} =\mu_{+}(t_2,U) -\mu(t_{1},U,n_1=2).
\label{deltacr}
\end{equation}
For a fixed ratio $U/t_1$, $\Delta_{c,r}$ behaves as
shown in Fig.\ \ref{tgap}. Above $n_c$, the gapless insulating state
I$_4$ is again stabilized. It should also be noted that both gaps 
($\Delta_r^*$ and $\Delta_{c,r}$) display universal behavior in the sense
that they do not depend on $\ell$. Note also that Eqs.\ (\ref{rstar}), 
(\ref{rc}), and (\ref{ri}) do not depend on $\ell$, so that $\overline{U}$
is also universal.

We now consider $U>\overline{U}$, an example of which is shown in Fig.\
\ref{pht}(b). For $r<r_c<r^*$, one finds the same sequence
I$_3$-M$_3$-I$_4$, with all insulating phases being gapless. If
$r_c<r<r^*$, a gapped insulating phase is crossed at
$n=n_c$. Similarly, for $r>r^*$ one goes from a metallic to a gapless
insulating phase (M$_2 \rightarrow$ I$_3$), and again crossing the
Mott-Hubbard phase at $n_c$.

The effective charge and spin velocities are given by 
\begin{equation}
\label{vets}
c_{\nu }=\frac{u_{1,\nu }(1+\ell )}{\sqrt{1+\Delta _{\nu }\ell ru_{1,\nu
}/u_{2,\nu }+\left( r\ell u_{1,\nu }/u_{2,\nu }\right) ^{2}}},
\end{equation}
which vanish for \( n<n^{\ast } \) and are smaller
than the velocities of the homogeneous system (\( c_{\nu }<u_{\nu } \)).
Furthermore, \( c_{\rho } \) displays re-entrant behavior as a function
of \( n \).

Finally, the effective interaction parameter \( K_{\rho }^{*} \)
is \begin{equation}
\label{krhot}
K_{\rho }^{\ast }=\frac{\sqrt{1+\Delta _{\nu }\ell ru_{1,\nu }/u_{2,\nu }+\left( r\ell u_{1,\nu }/u_{2,\nu }\right) ^{2}}}{\frac{1}{K_{1,\nu }}+\ell \frac{1}{K_{2,\nu }}r\frac{u_{1,\nu }}{u_{2,\nu }}}.
\end{equation}

\begin{figure}
{\centering \resizebox*{3.in}{!}{\includegraphics*{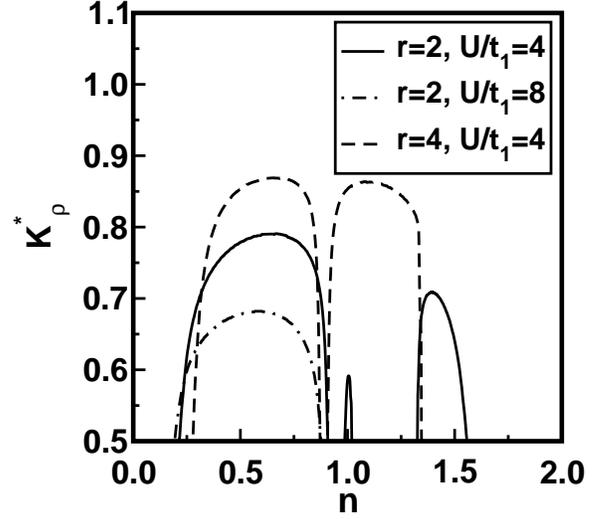}} \par}
\caption{\label{Kt}The effective correlation exponent \protect\( K_{\rho }^{*}\protect \)
(\protect\( \ell =1\protect \)) as a function of the band filling
\protect\( n\protect \) for different values of \protect\( U/t_1\protect \)
and \protect\( r\protect \).}
\end{figure}

In Fig.~\ref{Kt}, \( K_{\rho }^{\ast } \) is shown as a function of
band filling, for different couplings in superlattices with \( \ell =1
\): HSL-A with \( U=4t_{1} \) and \( r=2 \); HSL-B with \( U=8t_{1} \)
and \( r=2 \); HSL-C with \( U=4t_{1} \) and \( r=4 \). Note that \(
1/2<K_{\rho }^{\ast }<1 \) for any \( \ell \) in the metallic phases.
The various cases depicted in Fig.~\ref{Kt} show three (A), one (B)
and two (C) metallic phases.  In the homogeneous Hubbard chain, the
density-density and spin-spin correlation functions decay faster when
the hopping increases, since \(K_\rho\) increases with the ratio
\(t/U\). The effective correlation exponent of HSL-C is larger
than in HSL-A (see Fig.~\ref{Kt}), because of the larger hopping
amplitude of sub-chain 2 in HSL-C and the `averaging' nature of
\(K_\rho^\ast\).

We should stress that in the homogeneous system, the Luttinger Liquid
description breaks down at half-filling, when a gap opens in the
charge (though not in the spin) sector. In the superlattice, this
breakdown occurs in the insulating phases, as a result either of
Umklapp processes (Mott gap, lower phase of Fig.~\ref{hslfree}, phases
I$_1$ and I$_2$ of Fig.~\ref{tph16} and phases I$_2$ and I$_3$ of
Fig.~\ref{pht}), or of a band in one of the sub-lattices becoming
completely full or empty (upper phase of Fig.~\ref{hslfree}, phase
I$_3$ of Fig.~\ref{tph16} and phases I$_1$ and I$_4$ of
Fig.~\ref{pht}),

\section{CONCLUSIONS}

\label{conclusions}
We have discussed in full generality the properties
of Luttinger liquid superlattices. We have seen how most features
of a conventional Luttinger liquid description survive in the superlattice
structure. In particular, a few effective parameters, the spin and
charge velocities (\( c_{\rho } \) and \( c_{\sigma } \)) and the
stiffnesses (\( K_{\nu }^{\ast } \) and \( \overline{K}_{\nu } \))
are all that is required for a complete description of the low-energy
sector. These turn out to be combinations of the LL parameters of
the superlattice sub-units combined in proportion to their spatial
extent. As we have stressed in the Introduction, this opens the way
for possible `engineering' of Luttinger liquids.

This framework was applied to the study of the general phase diagram of
Hubbard superlattices. It was then illustrated how one can tune between
different phases by an appropriate choice of superlattice modulation. It
was found that the superlattice displays a variety of metallic and
insulating phases, the most prominent feature being the appearance of
gapless insulating phases, as a result of the one-dimensional character of
the system; gapped insulating phases were also found at some special
densities.

Single-wall metallic carbon nanotubes (SWMN's) seem to provide a
promising route towards realizing these LLSL's. Indeed,
notwithstanding the fact that SWMN's are, in general, described by a
less simplistic model (possibly even with more
branches\cite{eggeretal,eggergogolin1}), the LL coupling constant
depends on its (true) aspect ratio through\cite{eggeretal}

\begin{equation}
K_\rho = \left\{1+\frac{8e^2}{\pi\kappa\hbar v_F} \ln
\frac{L}{2\pi R} \right\}^{-1/2},
\label{g}
\end{equation}
where $\kappa$ is the dielectric constant, and $L$ and $R$ are,
respectively, the nanotube length and radius; typically one has
$K_\rho\simeq 0.2$-0.3.  More recently, the growth of intramolecular
junctions of SWMN's with different radii has been achieved with the
introduction of a pentagon and a heptagon into the hexagonal carbon
lattice,\cite{chico, collinsetal, marteletal, tansetal2, Yaoetal,
  kilicetal, andriotisetal} so that the fabrication of a superlattice
made up of SWMN's with different coupling constants has become a
concrete possibility.

We therefore expect the phase diagram of this `nanotube array' to
share several features with the general Hubbard superlattice. This is
because the only ingredients that enter into the phase determination
are the thermodynamic equilibrium condition and charge conservation.
In the case of a Luttinger liquid these can be easily written down if
one knows how the LL parameters depend on the density\[ \delta \mu
=\frac{\pi }{2L}\frac{v_{\rho }}{K_{\rho }}\delta N.\] Thus, the
sequence of insulating and metallic phases that we have found in
Hubbard superlattices should be present in other systems as well,
\emph{as they will reflect the phase diagram of the sub-units}. We
hope this rich variety of behaviors will stimulate further
experimental work along the lines of carefully controlled nanotube
arrays.

\begin{acknowledgments}
The authors are grateful to A.O.\ Caldeira, A.L.\ Malvezzi, and T.\
Paiva for discussions. Financial support from the Brazilian Agencies CNPq,
FAPESP (E.M.), and FAPERJ (R.R.d.S.) is also gratefully acknowledged.
\end{acknowledgments}
\appendix*

\section{Weak coupling bosonization of a Hubbard Superlattice}

\label{appendixA} Here we consider a Hubbard superlattice in weak
coupling and show that it is possible to describe the low-energy
properties
in terms of a Luttinger liquid superlattice. The Hamiltonian of a
Hubbard superlattice is \begin{eqnarray}
H & = & -t\sum _{j,\sigma }(\Psi _{j,\sigma }^{\dagger }\Psi _{j+1,\sigma }+h.c.)+\sum _{j}U_{j}n_{j,\uparrow }n_{j,\downarrow }\nonumber \label{hhc} \\
 & - & \mu \sum _{j,\sigma }\Psi ^{\dagger }_{j,\sigma }\Psi _{j,\sigma }.
\end{eqnarray}
We focus on the low energy modes near the Fermi surface, so that
each fermion is written as\cite{voit}
\begin{equation}
\label{psa}
\Psi_{j,\sigma }\approx e^{-ik_{F}ja}\Psi _{-,j,\sigma }+e^{ik_{F}ja}\Psi
_{+,j,\sigma },
\end{equation}
 where \( a \) is the lattice parameter, and the subscripts $+$ and
$-$ respectively denote right and left movers. The kinetic energy part
is then linearized as in the homogeneous case 
{\small 
\begin{eqnarray}
H_{0} & = & -t\sum _{j,\sigma }(\Psi _{j,\sigma }^{\dagger }\Psi _{j+1,\sigma }+h.c.)-\mu \sum _{j,\sigma }\Psi ^{\dagger }_{j,\sigma }\Psi _{j,\sigma }\nonumber \\
 & \approx  & v_{F}\sum _{\sigma }\int dx\Bigl [\Psi ^{\dagger }_{-,\sigma
}(x)\partial _{x}\Psi _{-,\sigma }(x)-\Psi ^{\dagger }_{+,\sigma
}(x)\partial _{x}\Psi _{+,\sigma }(x)\Bigr ].\nonumber \\ 
 & & 
\end{eqnarray}
} 

The fermionic fields are given in terms of the bosonic ones,
$\Phi_{\pm,\sigma}$, as\cite{voit}

\[
\Psi _{\pm ,\sigma }(x)=\frac{1}{\sqrt{2\pi \alpha }}U^{\dagger }_{\pm ,\sigma }e^{\mp 2i\sqrt{\pi }\Phi _{\pm ,\sigma }},
\]

Here $\alpha$ is a cutoff parameter and $U_{\pm,\sigma}$ is the Klein
factor.\cite{haldaneLL,voit}
Thus we get 
\begin{equation}
H_{0}=\frac{v_{F}}{2\pi }\sum _{\nu =\rho ,\sigma }\int dx\Bigl [(\partial _{x}\Theta _{\nu })^{2}+(\partial _{x}\Phi _{\nu })^{2}\Bigr ],
\end{equation}
where \( \Phi _{\nu}=(\Phi _{\uparrow}\pm \Phi _{\downarrow})/\sqrt{2} \)
and 
\begin{equation}
\Theta_{\nu }=\frac{1}{\sqrt{2}}[(\Phi_{+,\uparrow }-\Phi_{-,\uparrow })\pm (\Phi_{+,\downarrow }-\Phi_{-,\downarrow })].
\end{equation}

We now work out the low energy part of the on-site Hubbard
interaction. Again, we use Eq.(\ref{psa}) to get 
\begin{eqnarray}
H_{int} & = & \sum _{j}U_{j}:n_{j,\uparrow }::n_{j,\downarrow }:\nonumber \\
 & \approx  & \sum _{j}U_{j}\Bigl [(J_{+,j,\uparrow }+J_{-,j,\uparrow })(J_{+,j,\uparrow }+J_{-,j,\uparrow })\nonumber \\
 & + & (\Psi ^{\dagger }_{+,j,\uparrow }\Psi _{-,j,\uparrow }\Psi ^{\dagger }_{-,j,\downarrow }\Psi _{+,j,\downarrow }+h.c.)\Bigr ],
\end{eqnarray}
where 
$:\ldots:$ denotes normal ordering,\cite{voit}
\( J_{\pm ,j,\sigma }=:\Psi ^{\dagger }_{\pm ,j,\sigma }\Psi _{\pm
,j,\sigma }:=\frac{1}{\sqrt{\pi }}\partial _{x}\Phi _{\pm ,\sigma } \),
and the Umklapp terms have been neglected. Then 
\begin{eqnarray}
H_{int} & \approx  & a\int dx\ U(x)\frac{1}{\pi }(\partial _{x}\Phi
_{\uparrow })(\partial _{x}\Phi _{\downarrow })\nonumber \\
 & + & a\int dx\ U(x)\left[\frac{1}{(2\pi \alpha )^{2}}e^{2i\sqrt{\pi
}(\Phi _{\uparrow }-\Phi _{\downarrow })}+\text{h.c.}\right],\nonumber\\
& &
\end{eqnarray}
where h.c. stands for hermitian conjugate.
 In terms of charge (\( \rho  \)) and spin (\( \sigma  \)) fields
we have \begin{eqnarray}
H_{int} & \approx  & a\int \frac{dx}{\pi }U(x)\Bigl [(\partial _{x}\Phi _{\rho })^{2}+(\partial _{x}\Phi _{\sigma })^{2}\Bigr ]\nonumber \\
 & + & a\int dxU(x)\frac{1}{2(\pi \alpha )^{2}}\cos (\sqrt{8\pi }\Phi _{\sigma }),
\end{eqnarray}
 the last term corresponding to the spin backscattering interaction,
which is irrelevant in the RG sense. Finally, the low energy Hamiltonian
for the Hubbard superlattice is 
\begin{eqnarray}
H & = & \frac{v_{F}a}{2\pi }\int dx\left\{ (\partial _{x}\Theta _{\rho
})^{2}+\left[1+\frac{U(x)}{\pi v_{F}}\right](\partial _{x}\Phi _{\rho
})^{2}\, \right\} \nonumber \\
 & + & \frac{v_{F}a}{2\pi }\int dx\left\{ (\partial _{x}\Theta _{\sigma
})^{2}+\left[1-\frac{U(x)}{\pi v_{F}}\right](\partial _{x}\Phi _{\sigma
})^{2}\, \right\}.\nonumber\\ 
& &
\end{eqnarray}
This has the same form as Eq.~(\ref{llsl}), which describes the Luttinger
liquid superlattice.


\end{document}